\documentclass[fleqn,10pt]{wlscirep}
\usepackage[utf8]{inputenc}
\usepackage[T1]{fontenc}
\usepackage{graphicx}
\usepackage{caption}
\usepackage{subcaption} 
\usepackage{comment}
\usepackage{multirow}
\usepackage{float}
\usepackage{pdfpages} 
\usepackage{amsmath,siunitx}
\DeclareSIUnit\atom{atom}
\DeclareSIUnit\angstrom{\text{\AA}}
\DeclareSIUnit\bar{bar}

\usepackage{xr}
\makeatletter
\newcommand*{\addFileDependency}[1]{%
  \typeout{(#1)}%
  \@addtofilelist{#1}%
  \IfFileExists{#1}{}{\typeout{No file #1.}}%
}
\makeatother
\newcommand*{\myexternaldocument}[2][]{%
  \externaldocument[#1]{#2}%
  \addFileDependency{#2.tex}%
  \addFileDependency{#2.aux}%
}
\myexternaldocument[SI-]{Supplementary}
\newcommand{\Nrep}{eight}
\newcommand{\Natoms}{6000}

\newcommand{\CpDP}{24.9}                   
\newcommand{\CpGlassSlope}{26.1}
\newcommand{\CpLiqSlope}{39.5}
\newcommand{\dCpSlope}{13.4}

\newcommand{\dCpFluct}{14.1}
\newcommand{\GlassEdge}{800}               
\newcommand{\LiquidEdge}{1100}             

\newcommand{\TgSlope}{862}                 
\newcommand{\TgU}{905}                     
\newcommand{\TgUerr}{5}                    
\newcommand{\TgV}{918}                     
\newcommand{\TgExpA}{701}                  
\newcommand{\TgExpB}{706}                  
\newcommand{\TgRatioLo}{1.2}
\newcommand{\TgRatioHi}{1.3}

\newcommand{\AlphaGlass}{7.4}
\newcommand{\AlphaLiquid}{12.7}

\newcommand{\EqGlassLo}{15}                
\newcommand{\EqGlassHi}{22}
\newcommand{\EqHiLo}{130}                  
\newcommand{\EqHiHi}{150}

\newcommand{\RateSlow}{0.6}                
\newcommand{\RateStair}{0.16}              
\newcommand{\RampPerHundred}{167}          

\newcommand{\CpGlassRampA}{17.1}           
\newcommand{\CpGlassRampB}{20.0}           
\newcommand{\CpGlassRampC}{23.9}           
\newcommand{\CpGlassRampD}{25.2}           
\newcommand{\CpGlassStep}{25.7}            
\newcommand{\CpNineStep}{30.4}             
\newcommand{\CpNineRamp}{27.4}             
\newcommand{\CpKiloStep}{38.0}             
\newcommand{\CpKiloRamp}{32.9}             

\newcommand{\PotDiffTyp}{0.3}              
\newcommand{\PotDiffMax}{3.0}              

\newcommand{\PotA}{MACE-OMAT-LJ}
\newcommand{\PotB}{MACE-OMAT}
 
\usepackage{xcolor}

\title{Foundation interatomic potentials capture structural disorder in metallic glasses}
\author[1]{Dario Massa}
\author[2,+,*]{Luca Leoni}
\author[3,4,+]{Antoni Wadowski}
\author[1,7]{Kacper Cybinski}
\author[4]{Anshul D. S. Parmar}
\author[6]{Carmine Autieri}
\author[2,5]{Cesare Franchini}
\author[3]{Jan Wrobel}
\author[4,*]{Silvia Bonfanti}
\author[1,8,*]{Piotr Sankowski}

\affil[+]{These authors contributed equally to this work}
\affil[*]{Corresponding authors: piotr.sankowski@ideas.edu.pl, luca.leoni12@unibo.it, silvia.bonfanti@ncbj.gov.pl}

\affil[1]{IDEAS Research Institute, 13 Krakowskie Przedmieście Street, 00-071 Warsaw, Poland}
\affil[2]{Department of Physics and Astronomy 'Augusto Righi', Alma Mater Studiorum - Università di Bologna, Viale Berti Pichat 6/2, 40127, Bologna, Italy, Europe}
\affil[3]{Warsaw University of Technology, Faculty of Materials Science and Engineering, Wołoska St. 141, 02-507 Warsaw, Poland}
\affil[4]{NOMATEN Centre of Excellence, National Center for Nuclear Research, 05-400, Otwock, Poland, Europe}
\affil[5]{University of Vienna, Vienna Doctoral School in Physics, Boltzmanngasse 5, 1090, Vienna, Austria}
\affil[6]{International Research Centre Magtop, Institute of Physics,
Polish Academy of Sciences, Aleja Lotnik\'ow 32/46, PL-02668 Warsaw, Poland}
\affil[7]{University of Warsaw, Faculty of Physics, Ludwika Pasteura 5, 02-093 Warsaw, Poland}
\affil[8]{University of Warsaw, Faculty of Mathematics, Informatics, and Mechanics, Stefana Banacha 2, 02-097 Warsaw, Poland}

\graphicspath{{manuscript/}{Risposta/}}

\newsavebox{\panelbox}
\newcommand{\panelgraphic}[3]{%
  \sbox{\panelbox}{\includegraphics[width=#2]{#3}}%
  \usebox{\panelbox}%
  \llap{\raisebox{\dimexpr\ht\panelbox-2.4ex\relax}[0pt][0pt]{%
    \makebox[\wd\panelbox][l]{\hspace{0.8em}%
      \scalebox{0.85}{\sffamily\bfseries\large #1}}}}%
}
\newcommand{\panellabelfont}{\captionfont\captionlabelfont}
\newcommand{\figlabel}[3]{%
  \llap{\raisebox{\dimexpr\ht\panelbox-#2\ht\panelbox\relax}[0pt][0pt]{%
    \makebox[\wd\panelbox][l]{\hspace*{#1\wd\panelbox}{\panellabelfont #3}}}}}

\begin{abstract} 

Foundation machine-learning interatomic potentials have recently shown remarkable transferability across broad chemical and structural spaces, yet their reliability for highly disordered metallic states remains less established. Metallic glasses (MGs) provide a particularly demanding benchmark: they have traditionally been modeled with empirical Embedded Atom Method (EAM) potentials, which are fitted to specific target compositions and have limited transferability outside their fitting domain, while \textit{ab initio} molecular dynamics (AIMD) remains prohibitively expensive at the large system sizes and lengthy timescales required for amorphous metals. Here we demonstrate that foundation potentials are ready to describe structural disorder---reproducing the structure, quenching dynamics, and mechanical and thermal properties of the amorphous state at near first-principles accuracy---using the prototypical Cu-Zr-Al MG as a rigorous benchmark. 

We systematically evaluate the performance of MACE foundation models against a classical EAM potential by analyzing Molecular Dynamics (MD) quenching trajectories, validating structural, elastic, mechanical, and thermal properties against Density Functional Theory (DFT) and experimental baselines. Focusing initially on a reference composition ($\text{Cu}_{0.46}\text{Zr}_{0.46}\text{Al}_{0.08}$), we then extend our evaluation across a varied compositional space within the expected good glass-forming range to assess the models' generalization power. Furthermore, we investigate the efficacy of specializing these potentials using two distinct data-generation workflows: physically-guided thermodynamic sampling via Lennard-Jones (LJ) surrogate models, and automated, symmetry-driven structural exploration via small periodic cells (ASSYST).

Our results demonstrate that foundation models pre-trained on expansive databases containing non-equilibrium liquid and highly amorphous states---such as the OMAT database---already achieve high predictive accuracy and compositional generalization out-of-the-box, significantly outperforming the composition-specific EAM potential. At the same time, fine-tuning on target MG structures derived from LJ surrogate models yields only modest improvements over the pre-trained baseline, whereas specialization on the automated small-cell database (ASSYST) proves detrimental for capturing disordered dynamics. Taken together, these findings indicate that the readiness of foundation potentials for disorder is set primarily by the diversity of their pre-training data rather than by system-specific fine-tuning, shifting the modeling effort from bespoke dataset generation toward the coverage of pre-training corpora and the careful validation of the resulting dynamics.

\end{abstract}
\begin{document}

\flushbottom
\maketitle

\section{Introduction}

Machine-learning interatomic potentials (MLIPs) have transformed atomistic modeling by mapping the high-dimensional configuration of a material directly onto its quantum-mechanical potential energy surface (PES), from which atomic forces are obtained as the analytical gradient of the predicted energy\cite{park2026machine}. Trained on high-fidelity Density Functional Theory (DFT) data, they capture intricate many-body interactions within the classical Molecular Dynamics (MD) framework. Their most recent evolution is that of foundation MLIPs\cite{riebesell2025framework,mazitov2025pet,rhodes2025orb,lysogorskiy2026graph,batatia2022mace,tahmasbi2026benchmarking,liu2025heterogeneous,kong2026scalable}: pre-trained on extremely large and diverse datasets\cite{barroso2024open,Deng2023}, they aim to act as zero-shot force evaluators across nearly the entire periodic table.

These models are, however, intrinsically bound by their training distributions and often struggle to extrapolate to out-of-distribution scenarios. The amorphous state is emerging as a central challenge in this respect, since early pre-trained models and benchmarks have largely targeted ordered, crystalline structures, and their transferability to non-crystalline solids is far from established\cite{fragapane2026amorphous}. The difficulty is compounded by the fact that accuracy on single-point energies and forces does not necessarily translate into a faithful description of structural and dynamical observables at finite temperature, as recent finite-temperature benchmarks such as Dyna-Mat have highlighted\cite{gawkowski2026dyna}. To mitigate this distributional gap, fine-tuning strategies are increasingly adopted to specialize pre-trained models\cite{liu2026fine,radova2025fine,Angeletti2025}.

Metallic glasses (MGs)\cite{klement1960non,bonfanti2026} provide an especially demanding test of such transferability. Lacking long-range crystalline periodicity and microstructural defects such as grain boundaries and dislocations, they combine low elastic moduli with high yield strengths, large elastic limits and good thermoplastic formability\cite{wang2009bulk}; more importantly for the present purpose, their disordered structure probes exactly the local chemical environments that a general-purpose potential must learn rather than memorize. Traditionally, MGs have been modeled with classical MD driven by Embedded Atom Method (EAM) potentials\cite{daw1984embedded}; the Mendelev Cu-Zr potential\cite{mendelev2009development}, for instance, was parameterized against experimental diffraction data and first-principles liquid mixing enthalpies, reproducing the liquid-to-glass transition temperature in good agreement with calorimetry. Such empirical potentials, however, have limited transferability outside their parameterization domain and are ill-suited to explore the wide compositional space of MGs, while \textit{ab initio} molecular dynamics (AIMD) remains prohibitively expensive at the system sizes and timescales required for amorphous metals.

For the specific problem of disorder, the composition of the pre-training corpus is decisive: databases that explicitly include high-temperature dynamics and non-equilibrium amorphous states, such as the Open Materials (OMAT) database\cite{barroso2024open}, offer a promising route to model MGs directly. Building a representative training set for these systems is itself non-trivial. Modern workflows integrate active learning with advanced thermodynamic sampling---using Lennard-Jones (LJ) surrogate models combined with swap Monte Carlo (swap-MC) moves to computationally age the glass toward deeply relaxed states\cite{wadowski2026efficient}---but require DFT on large supercells and must be repeated for each composition of interest. Alternatively, automated pipelines built on small periodic cells, such as ASSYST\cite{poul2025automated}, yield reliable potentials for the energetics of complex crystalline systems, although their ability to capture the dynamics of structurally disordered phases is less clear.

In this work, we focus on the widely studied Cu-Zr-Al metallic glass, which enables comparison with a body of computational and experimental studies\cite{jiang2024situ,poltronieri2023mechanical,makinen2025avalanches,inoue2004formation,cheung2007thermal,fan2006thermophysical,yu2008poisson,idriss2015evolution,li2009effect,plummer2009elastic}. Complementary to broad, multi-system surveys of the amorphous challenge\cite{fragapane2026amorphous}, we provide an in-depth study of a glass-forming family: we assess the capability of MACE foundation potentials\cite{batatia2022mace} to reproduce the quenching dynamics and the structural, elastic, mechanical and thermal properties of a reference stoichiometry, as well as their extrapolation across composition. We compare pre-trained MACE models with fine-tuned versions specialized on i) an LJ-surrogate database of representative quenched configurations, and ii) an ASSYST database of small cells subject to random rattling of atomic positions. Together, these comparisons address the practical applicability of general-purpose potentials to MGs---based both on their pre-training databases and on modern data-generation workflows---delivering an overview from composition-specific to composition-dependent properties.


\section{Results}
This Section assesses how foundation potentials describe the Cu-Zr-Al MG, moving from a single reference stoichiometry, $\text{Cu}_{0.46}\text{Zr}_{0.46}\text{Al}_{0.08}$, to the whole good glass-forming ability (GFA) region of the ternary.

The comparison involves a classical EAM potential\cite{mendelev2009development} as baseline, four pre-trained MACE foundation models that differ essentially in their pre-training corpus---OMAT\cite{barroso2024open}, which explicitly contains high-temperature and off-equilibrium configurations, against the near-equilibrium MPTrj\cite{Deng2023} and MPTrj+sAlex\cite{schmidt2023machine} data---and two fine-tuned versions of the MACE-OMAT backbone, hereafter MACE-OMAT-LJ and MACE-OMAT-ASSYST, specialized respectively on: i) a Lennard-Jones (LJ) surrogate database of quenched target-composition structures; and ii) an automatically generated small-cell database (ASSYST). Both data-generation strategies and the fine-tuning protocol are detailed in the Methods (Sec.~\ref{sec:Data generation strategies} and Sec.~\ref{sec:Training strategy}), with dataset and hyperparameter specifics reported in the Supplementary Material.

Across all potentials we benchmark the thermodynamic quantities sampled along MD quenching (energies, atomic forces and cell stresses, against static DFT), the short-range order encoded in the radial distribution functions, the isotropic elastic constants ($E$, $B$, $G$, $\nu$), the mechanical response under simple shear, and the constant-pressure heat capacity across the glass transition, following the simulation protocols of Sec.~\ref{sec:Molecular dynamics validation}.
Preliminary to any deployment in dynamics, we asked the question of how accurately a model reproduces energies, forces and stresses ($E$, $F$, $S$), monitored either zero-shot on held-out splits or along the fine-tuning protocol on a reference dataset. We report this stage in Methods: the original LJ database provides the arena for a first zero-shot screening among the pre-trained models (Sec.~\ref{sec:Screening strategy}), after which one of the best-scoring backbones is fine-tuned on the LJ and ASSYST databases following two distinct strategies, single-head full-parameter (Naive-FT) and multi-head with replay (MH-FT) (Sec.~\ref{sec:Training strategy}); hyper-parameter searches, learning curves, parity plots and per-model metrics are collected in Supplementary Secs.~\ref{SI-si:hp-lj},~\ref{SI-si:hp-as} and~\ref{SI-si:additional-results}. 
\subsection{Performance on reference Cu-Zr-Al system}\label{sec:Performance on reference CuZrAl system}
\subsubsection{MD quenching - benchmark}
\begin{figure}[htbp]
    \centering

    \begin{minipage}{0.99\textwidth}
        \centering
        \panelgraphic{a)}{\linewidth}{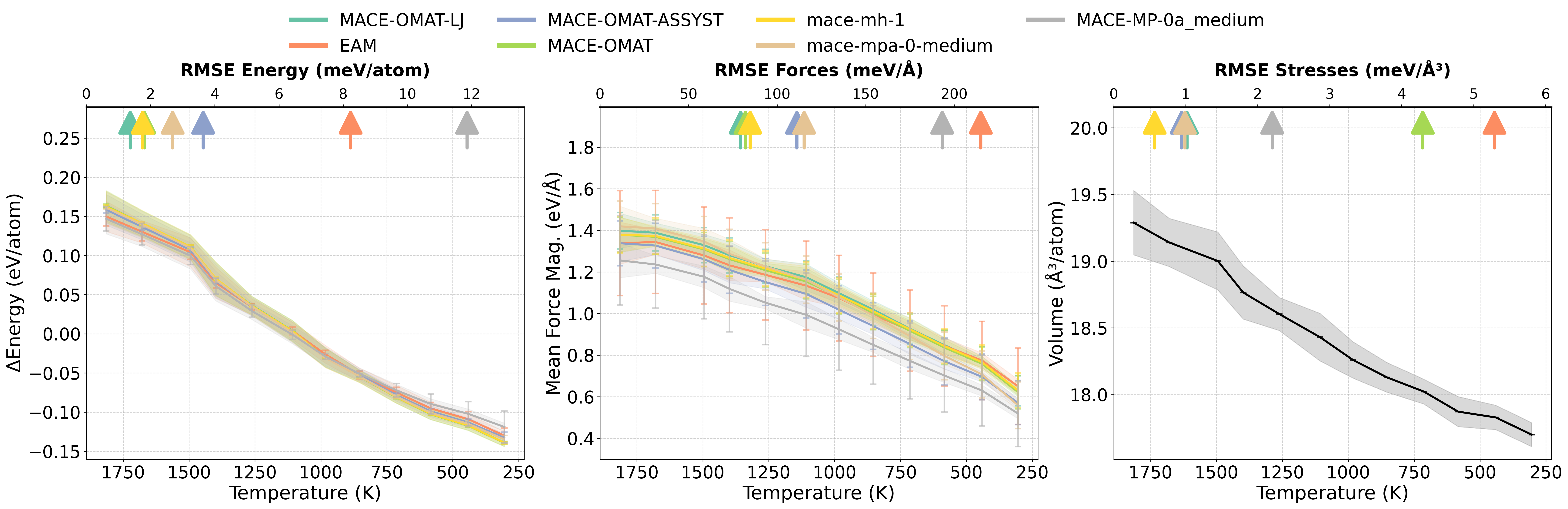}
    \end{minipage}

    \vspace{0.5ex}

    \begin{minipage}{0.99\textwidth}
        \centering
        \panelgraphic{b)}{\linewidth}{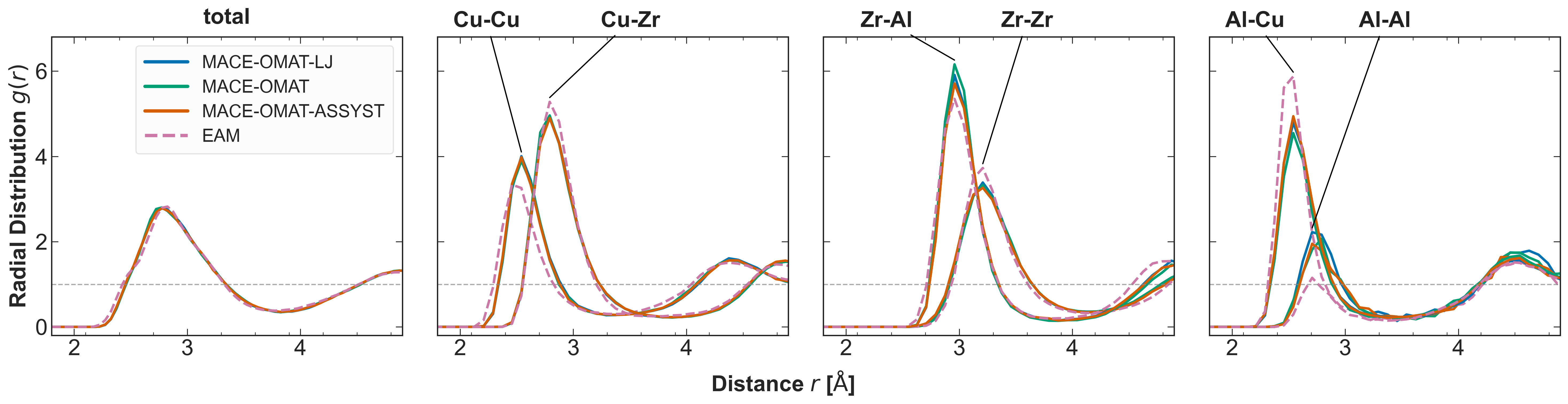}
    \end{minipage}

    \caption{Benchmark of the evaluated interatomic potentials on the reference $\text{Cu}_{0.46}\text{Zr}_{0.46}\text{Al}_{0.08}$ glass.
    \textbf{a)} Analysis of the quenching trajectories as generated from the different interatomic potentials and evaluated through DFT calculations. The panels track the mean energy, force magnitude, and volume as generated by various interatomic potentials and evaluated against DFT. To ensure statistical robustness, $17$ independent MD replicas were simulated for each potential, from which $12$ frames were extracted for static DFT evaluation. Solid colored lines track the mean predicted values, while the shaded regions indicate the $\pm 1$ standard deviation ($\sigma$) configurational variance across the replicas. Vertical error bars represent the Root Mean Square Error (RMSE) of the potentials' predictions relative to the DFT baseline at each specific temperature step. Global error scores, integrated over the entire quenching trajectory, are displayed on the secondary top horizontal axes, with colored arrows mapping each model to its cumulative performance. A single, neutral volume curve (black) is presented in the third panel; because the static evaluations are performed on fixed atomic coordinates, the simulation cell volume is an identical, intrinsic property of the extracted frames across all models.
    \textbf{b)} Total and partial radial distribution functions, $g(r)$, of the same system at 300 K, obtained from the same MD trajectories and averaged over ten independent runs. The panels display the total pair distribution alongside all individual atomic pairs (Cu-Cu, Cu-Zr, Cu-Al, Zr-Zr, Zr-Al, and Al-Al) across an interatomic distance range of $r = 1.8$ to $4.9\ \text{Å}$. Calculations are compared using different interatomic potentials: MACE-OMAT-LJ (blue), MACE-OMAT (teal), MACE-OMAT-ASSYST (orange), and a traditional embedded-atom method (EAM) potential (pink dashed line). The horizontal dashed line denotes the ideal disordered limit where $g(r) = 1$.
    }   
    \label{fig:trajectory_Target_analysis}
\end{figure}

To assess the readiness of foundation models for structural disorder---i.e.\ their ability to reproduce the first-principles potential-energy surface of the disordered state along its thermal evolution---we first evaluated their performance during the simulated MD quenching of the target composition. Figure~(\ref{fig:trajectory_Target_analysis}) tracks the thermodynamic evolution of the system, benchmarking the potentials' predictions of energy, atomic forces, and cell stresses against static DFT evaluations. 
The foundation models, particularly the pre-trained MACE-OMAT and its LJ-fine-tuned variant (MACE-OMAT-LJ), demonstrate exceptional predictive accuracy out-of-the-box. Both models maintain a minimal energy Root Mean Square Error (RMSE) of roughly $1-2$ meV/atom and a force RMSE of $\sim 80$ meV/Å across the entire temperature range. In stark contrast, the classical Embedded-Atom Method (EAM) potential exhibits larger energy offsets and a high force error exceeding $200$ meV/Å. Interestingly, fine-tuning the foundation model on the automated small-cell database (MACE-OMAT-ASSYST) degrades its energetic and force predictive capabilities compared to the pre-trained baseline, increasing the energy RMSE to $\sim 3.8$ meV/atom and force RMSE to $\sim 110$ meV/Å.

The structural representation of the amorphous phase is examined through the total and partial radial distribution functions, $g(r)$, extracted from the same MD trajectories and reported in Fig.~(\ref{fig:trajectory_Target_analysis}b). While the total $g(r)$ appears relatively consistent across all evaluated potentials, the partial distributions expose the differences in the EAM baseline: it slightly deviates from the MLIPs' representation of the local chemical ordering, by overestimating the first coordination peaks of Cu-Al and Zr-Zr, while underestimating Al-Al and Cu-Cu interactions. Conversely, all three MACE variants (pre-trained, LJ-fine-tuned, and ASSYST-fine-tuned) almost perfectly overlap across every atomic pair. Since no experimental structural reference is included here, this comparison demonstrates the mutual consistency of the MACE variants rather than an absolute validation: regardless of their specific fine-tuning, they learn a robust and unified representation of the complex, disordered local environments inherent to the MG.

\subsubsection{Elastic property -  benchmark}
\begin{figure} [t]
    \centering
    \includegraphics[width=0.99\textwidth]{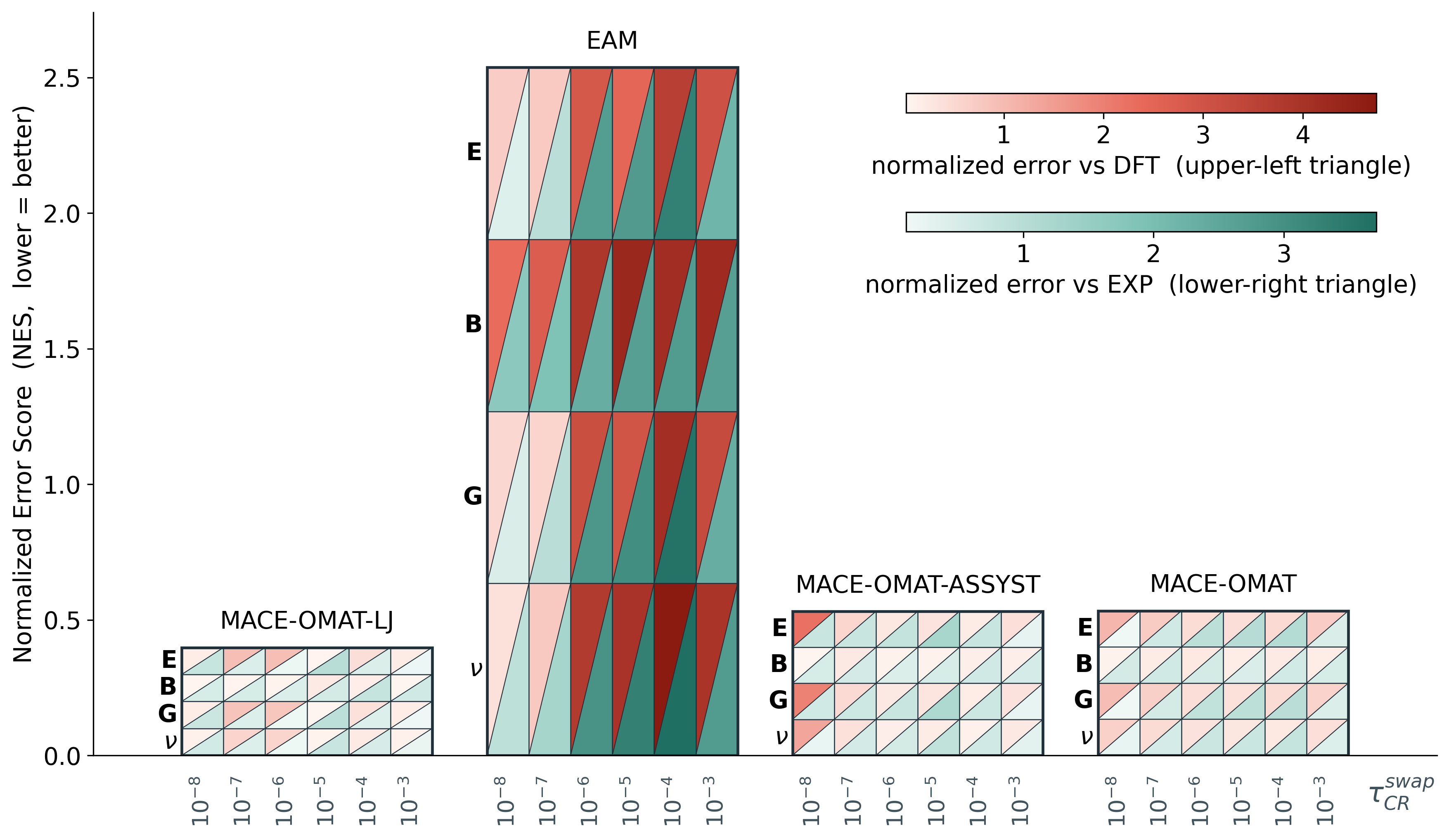}
    \caption{Aggregate benchmark of interatomic potentials on the elastic-property
    prediction task. For each potential we evaluate four elastic observables---Young's modulus $E$, bulk modulus $B$, shear modulus $G$, and Poisson's ratio $\nu$---on
    amorphous configurations generated at six swap-Monte-Carlo cooling rates $\tau_{\mathrm{CR}}^{\mathrm{swap}}\in[10^{-8},10^{-3}]$ expressed in the LJ reduced units, corresponding to quenching rates between $[10^{-20},10^{-8}]$ \SI{}{\kelvin\per\pico\second}, thanks to swap-Monte-Carlo acceleration \cite{wadowski2026efficient}. Then, we compare the predictions against two independent references: the corresponding DFT values (evaluated on the \emph{same} structures) and an experimental
    literature value. Each potential is shown as a single bar whose \emph{height} is its
    Normalized Error Score (NES; lower is better), as defined in Sec.~\ref{sec:Structural and elastic properties}. Within each bar, a $4\times 6$ grid resolves the underlying errors: rows are the four properties ($E,B,G,\nu$, top to bottom) and columns are the six cooling
    rates. Every cell is split along its diagonal into two triangles that encode the normalized error against DFT (upper-left, red colormap) and against experiment\cite{jiang2008zr} 
    (lower-right, teal colormap), with the two shared color scales given by the
    stacked colorbars. Supplementary Figs.~(\ref{SI-fig:supp_elastic_lj})--(\ref{SI-fig:supp_elastic_omat}) plot the actual predicted moduli against both references to resolve deviations for each potential. 
    }
    \label{fig:elastic_nes_grid}
\end{figure}

The elastic-property benchmark is summarized through the aggregate Normalized Error Score (NES), together with its property- and cooling-rate-resolved decomposition (Fig.~(\ref{fig:elastic_nes_grid})).
The three MACE-OMAT variants possess low NES of $0.40$–$0.53$, whereas the classical EAM potential reaches $2.54$—roughly five times larger. In absolute terms, the FPs predict the elastic moduli of the glass to within $2$–$3$~GPa of DFT on average, against $\approx17$~GPa for EAM. The data-driven potentials therefore reproduce the elastic response of these amorphous configurations at near-DFT fidelity, while the embedded-atom description does not. Within the MACE-OMAT family, MACE-OMAT-LJ is the most accurate with a NES of $0.40$; notably, the ASSYST fine-tuning leaves the elastic predictions essentially unchanged relative to the bare foundation model, which is reflected in the respective NES of $0.531$ vs. $0.533$.
The per-property rows reveal that, for all MACE potentials, the residual error is concentrated in the Young's ($E$) and shear ($G$) moduli, while the bulk modulus $B$ and Poisson's ratio $\nu$ are better captured (e.g.\ $\mathrm{RMSE}(B)<1.3$~GPa and $\mathrm{RMSE}(\nu)\sim0.01$ against DFT). EAM instead fails across all four observables and, more importantly, in a systematic and qualitative manner: it overestimates $E$ and $G$ while underestimating $B$ and $\nu$ (mean $E=134$, $G=54$, $B=88$~GPa and $\nu=0.24$, against the DFT values $109$, $41$, $114$~GPa and $0.34$). It thus describes a glass that is too stiff under shear yet too compliant under hydrostatic compression, with a low Poisson ratio—a distortion of the elastic character rather than a mere quantitative offset.
The sign and cooling-rate dependence of these deviations are resolved per potential in Supplementary Figs.~(\ref{SI-fig:supp_elastic_lj})--(\ref{SI-fig:supp_elastic_omat}), which plot the actual predicted moduli against both references. 

The cooling-rate axis exposes a further, physically informative trend. EAM potential's error is strongly rate-dependent, rising from near-quantitative agreement for the most slowly cooled, deeply supercooled samples to $3$–$4\times$ larger errors for the fast-quenched, more liquid-like glasses (see Supplementary Fig.~(\ref{SI-fig:supp_elastic_eam}) for a property- and rate-specific breakdown). This indicates that this potential captures the elastic response of well-relaxed, low-energy inherent structures but degrades markedly for the disordered, higher-energy configurations retained at fast quench.

Finally, comparing the two references (upper-left vs. lower-right triangles) shows that, for every machine-learned potential, the error against experiment systematically exceeds the error against DFT. This gap is not a deficiency of the potentials but a property of the reference electronic-structure method: DFT itself is offset from the measured values. In particular, first-principles predictions are over-stiff in $E$ and $G$, under-stiff in $B$, and low in $\nu$ with respect to the experimental values. Because the MACE-OMAT potentials faithfully reproduce their DFT training target, they inherit this offset, which therefore sets the floor on their achievable agreement with experiment.

\subsubsection{Shear-stress response}
\begin{figure} [t]
    \centering
\includegraphics[width=0.70\textwidth]{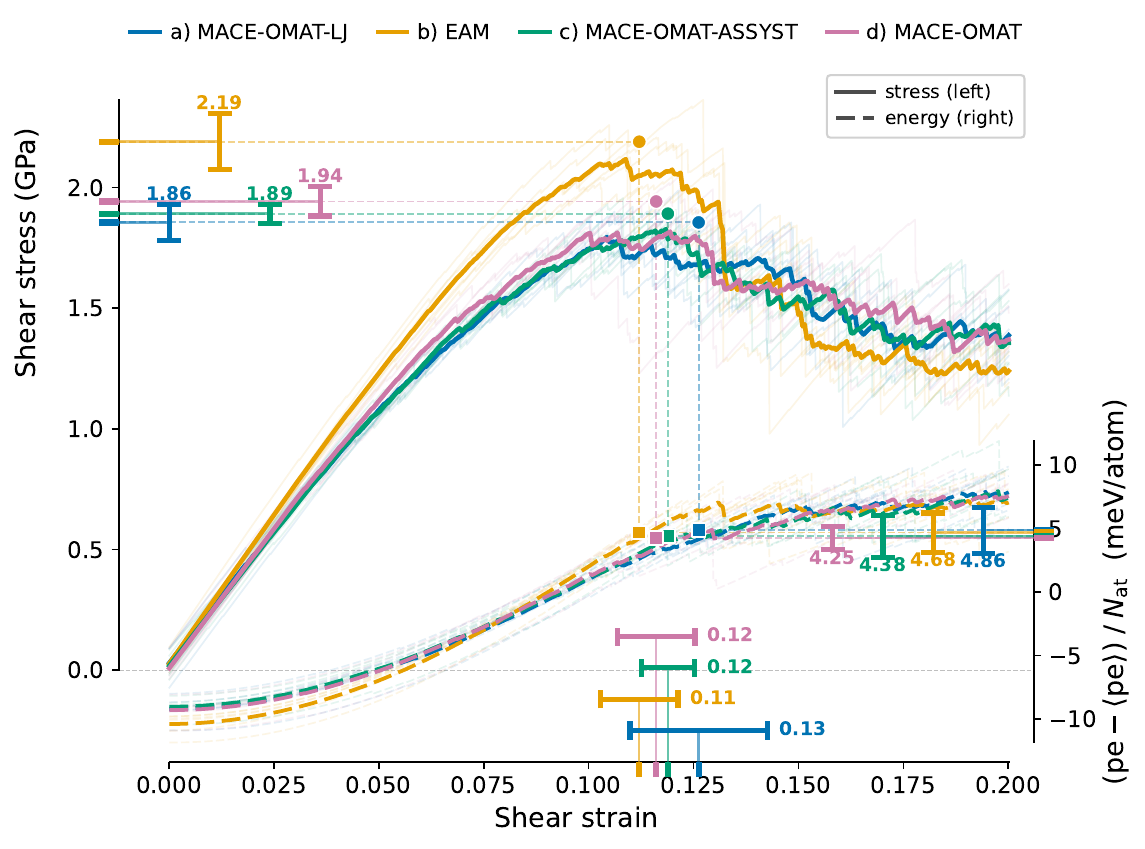}
    \caption{Shear stress (left) and shifted potential energy per atom (right) as a function of shear strain for four interatomic potentials on the $\text{Cu}_{0.46}\text{Zr}_{0.46}\text{Al}_{0.08}$ system. Each potential is represented by a distinct color (see legend); thin transparent lines are individual statistical realizations and the bold line is the mean response interpolated on a common strain grid. The right panel shows $(\mathrm{pe}-\langle\mathrm{pe}\rangle)/N_\mathrm{at}$ (meV/atom), where $\langle\mathrm{pe}\rangle$ is the grand mean of the potential energy over all strain steps and realizations for each potential. In both panels, colored dots mark the mean yield state, defined as the shear-stress maximum averaged over realizations; dashed crosshairs project this state onto both axes. Staggered floating error bars on each axis report the standard deviation ($\pm 1\sigma$); the thin colored stems connect each error bar to its reference position on the spine, and the numerical mean value is printed above the upper cap. Note that the annotated yield state in the above curves is evaluated at the strain of maximum stress — not at the maximum of the energy curve — so that both panels refer to the same physical event and can be read in conjunction. The combined panels share identical x-axis limits to facilitate direct comparison across potentials.
    }
    \label{fig:shear_combined}
\end{figure}

The shear stress and the excess potential energy per atom $(\mathrm{pe} - \langle \mathrm{pe}\rangle)/N_\mathrm{at}$, with $N_\mathrm{at}$ the number of atoms, are shown in Fig.~(\ref{fig:shear_combined}) (top and bottom panels, respectively) as a function of applied shear strain for all four interatomic potential descriptions, each sampled over six independent amorphous realizations.
All four potentials qualitatively reproduce the same characteristic mechanical response: a linear elastic rise at small strains, a sublinear hardening region, a distinct yield peak, and a discontinuous post-yield regime in which the stress drops in discrete events before settling into a lower-stress flowing state. The pre-yield regime is highly informative in our comparison. In this regime, the mean stress–strain curves are nearly superimposed across all foundation potentials up to approximately 50–60\% of the peak strain, indicating that the initial elastic response and early stages of plastic deformation are largely indistinguishable among these. The potentials diverge most visibly at and around the yield point. EAM predicts the highest ultimate shear strength $\tau_u$ and the earliest yield $\gamma_u$, while the three MACE-based potentials cluster in a significantly narrower band. The spread among the three MACE variants is only $\sim$3\%, which reflects the internal consistency of that family despite differences in training protocol. The EAM overestimate of $\tau_u$ by roughly 18\% relative to the MACE mean is notable and might lie in the reduced capability to capture higher-order effects contributing to the local resistance to shear, as well as in limited sampling for training (see Sec.~\ref{sec:EAM focus}). The run-to-run variability, reported as $\pm 1\sigma$ error bars on both axes, is comparable across potentials in the yield-stress direction ($\lesssim 5\%$), but MACE-OMAT-LJ shows a notably wider spread in yield strain relative to EAM, suggesting that the fine-tuning on specific configurations from the target MG system makes the onset of shear-band nucleation more sensitive to the specific initial amorphous configuration.

\subsubsection{Heat capacity and the glass transition}
\label{sec:results-cp}
To characterize the thermodynamic liquid-to-glass transition, we evaluated the specific heat
capacity ($C_p$) as a function of temperature during the simulated heating of the quenched structures. 
\begin{figure} [t]
    \centering
    \sbox{\panelbox}{\includegraphics[width=0.99\textwidth]{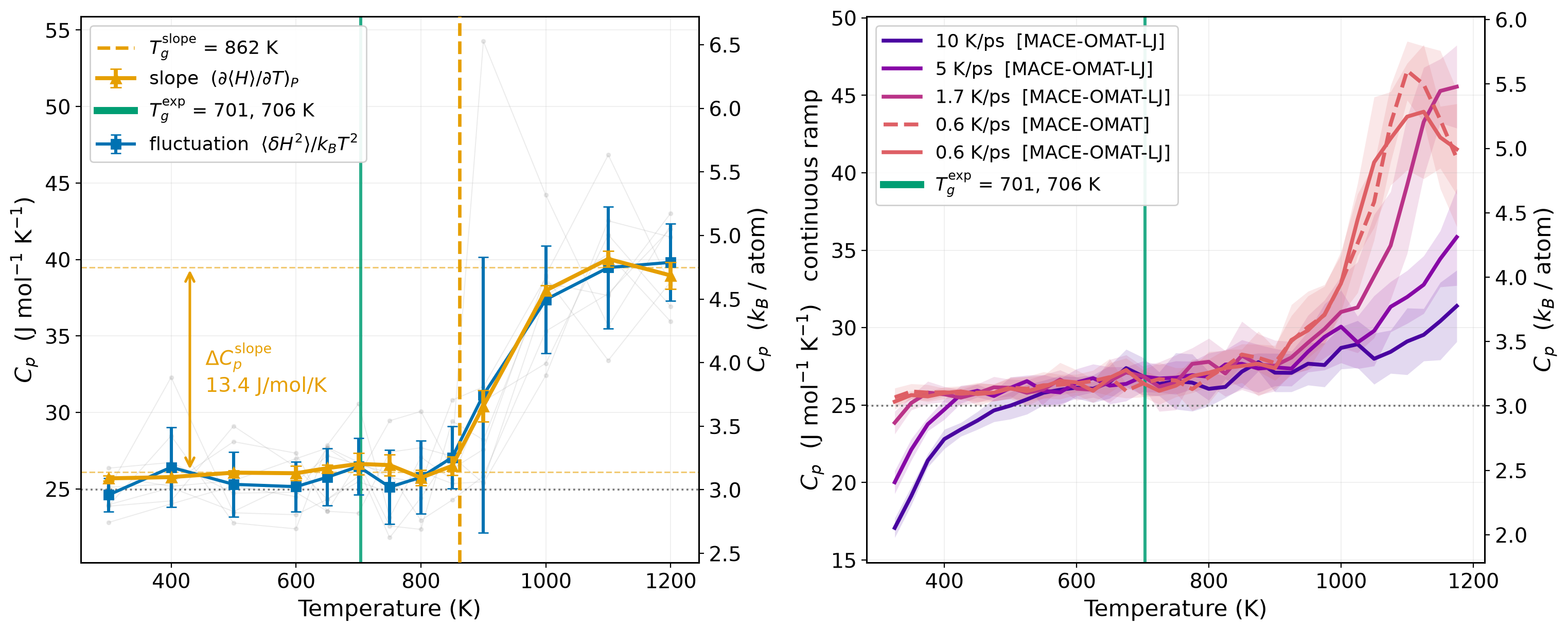}}%
    \usebox{\panelbox}%
    \figlabel{0.002}{0.06}{a)}%
    \figlabel{0.500}{0.06}{b)}%
    \caption{Constant-pressure heat capacity of Cu$_{46}$Zr$_{46}$Al$_{8}$ across the glass transition, and its dependence on heating rate. \textbf{(a)}~$C_p(T)$ obtained by stepwise heating of \Nrep{} independently quenched \num{\Natoms}-atom glasses with the \PotA{} potential, each temperature sampled under isothermal $NPT$ conditions. Blue squares: fluctuation route, $C_p=\langle\delta H^{2}\rangle/k_BT^{2}$, with the kinetic contribution treated analytically. Orange triangles: calorimetric slope route, $C_p=(\partial\langle H\rangle/\partial T)_P$, evaluated from plateau averages. Error bars denote $\pm 1$~s.d.\ across the \Nrep{} replicas; faint grey lines show the individual replicas. The dotted horizontal line marks the Dulong--Petit value of $\SI{\CpDP}{\joule\per\mole\per\kelvin}$. Dashed orange horizontals are the glass ($T\le\SI{\GlassEdge}{\kelvin}$) and supercooled-liquid ($T\ge\SI{\LiquidEdge}{\kelvin}$) baselines, whose separation defines the step $\Delta C_p^{\mathrm{slope}}=\SI{\dCpSlope}{\joule\per\mole\per\kelvin}$ (double arrow); the dashed orange vertical marks $T_g^{\mathrm{slope}}=\SI{\TgSlope}{\kelvin}$, taken where $C_p$ has traversed \SI{10}{\percent} of the step. The two estimators are obtained independently and are not adjusted to agree. \textbf{(b)}~$C_p(T)$ from \emph{single continuous} heating ramps at \SIlist{10;5;1.7;0.6}{\kelvin\per\pico\second}, obtained from the same equilibrated \SI{300}{\kelvin} configurations as in~(a) and analyzed solely through $\mathrm{d}H/\mathrm{d}T$ along the scan; shaded bands are $\pm 1$~s.d.\ over \Nrep{} replicas. Color encodes the heating rate (light to dark with decreasing rate) and line style the interatomic potential: solid, \PotA{}; dashed, \PotB{}.  The green vertical band in both panels indicates the experimental glass transition temperature (\SI{\TgExpA}{\kelvin}, Zhou \textit{et al.}; \SI{\TgExpB}{\kelvin}, Zhang \textit{et al.}) \cite{zhang2007glass,zhou2013microstructure}. Right-hand axes give the same quantity in $k_B$ per atom.}
    
    \label{fig:tg-overview}
\end{figure}
We use a stepwise protocol in which the samples are carried through a staircase of
isothermal plateaus joined by short ramps, so that every measurement is made at a
well-defined state point; this is what allows $C_p$ to be extracted from equilibrium enthalpy
fluctuations at each temperature, and it also makes a second, independent estimator---the
calorimetric slope $(\partial\langle H\rangle/\partial T)_P$ between successive
plateaus---available from the same trajectory, both described in detail in Sec.~\ref{sec:methods} and Supplementary Sec.~\ref{SI-si:cp-methods}.

Fig.~(\ref{fig:tg-overview}a) shows the resulting $C_p(T)$ for \Nrep{} independently quenched glasses. Below \SI{\GlassEdge}{\kelvin} the heat capacity is flat and sits at the Dulong--Petit value, $\SI{\CpGlassSlope}{\joule\per\mole\per\kelvin}$ against $3k_B=\SI{\CpDP}{\joule\per\mole\per\kelvin}$, as expected for a solid whose accessible excitations are purely vibrational. Above \SI{900}{\kelvin} it rises to a liquid plateau of $\SI{\CpLiqSlope}{\joule\per\mole\per\kelvin}$, the additional capacity reflecting the configurational degrees of freedom that become accessible once the structure is able to rearrange on the observation timescale. The step between the two branches, $\Delta C_p = 13.4 \pm 1.2 \,\, \text{J}\,\,\text{mol}^{-1}\text{K}^{-1}$  is the calorimetric signature of the glass transition, and its onset defines $T_g^{\mathrm{slope}}=\SI{\TgSlope}{\kelvin}$.

\begin{figure} [t]
    \centering
    \sbox{\panelbox}{\includegraphics[width=0.99\textwidth]{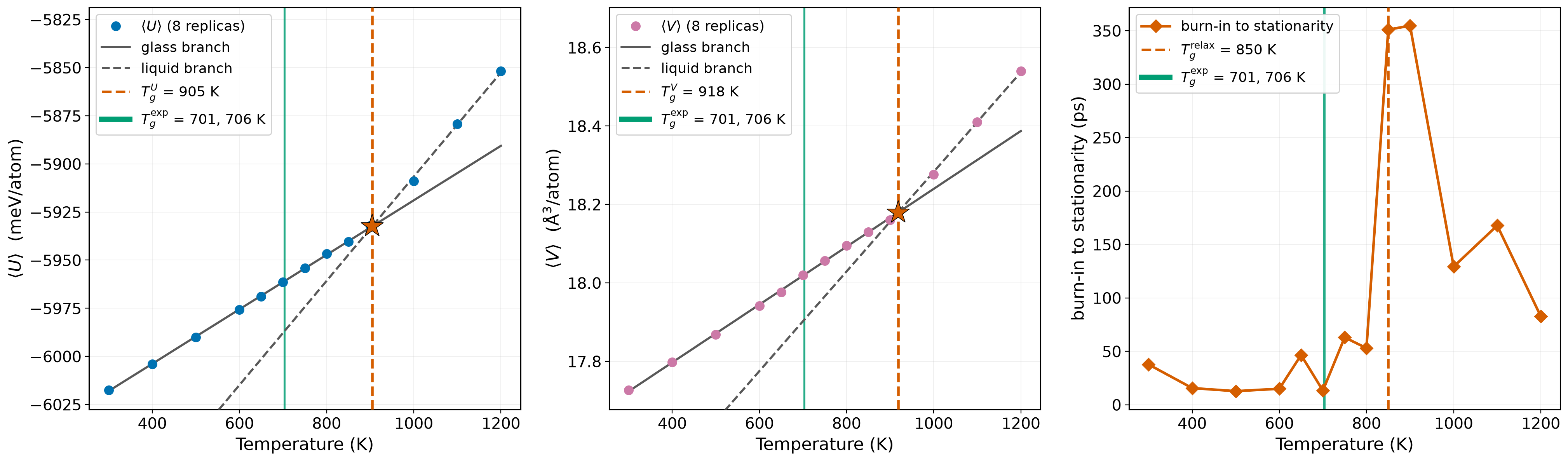}}%
    \usebox{\panelbox}%
    \figlabel{-0.012}{0.06}{a)}%
    \figlabel{0.352}{0.06}{b)}%
    \figlabel{0.670}{0.06}{c)}%
    \caption{\textbf{Three independent signatures of the glass transition from the same
simulations.}
\textbf{(a)} Mean potential energy per atom and \textbf{(b)} mean atomic volume, both
averaged over the eight replicas at each plateau. Solid and dashed grey lines are
least-squares fits to the glass and supercooled-liquid branches; stars mark their intersection, giving $T_g^{U}=\SI{905}{\kelvin}$ and
$T_g^{V}=\SI{918}{\kelvin}$ (dashed orange verticals). The thermal expansivity increases from
\SI{7.4e-4}{\cubic\angstrom\atom^{-1}\kelvin^{-1}} in the glass to
\SI{12.7e-4}{\cubic\angstrom\atom^{-1}\kelvin^{-1}} in the liquid, a factor of \num{1.7}.
Because $P_{\mathrm{ext}}V$ contributes only \SI{0.01}{\milli\electronvolt\per\atom} at
ambient pressure, $\langle U\rangle$ and $\langle H\rangle$ are numerically interchangeable
here and yield the same transition temperature; the volume therefore provides the independent
thermodynamic signature. \textbf{(c)} Time that must be discarded from the start of each
plateau before its enthalpy record satisfies a stationarity criterion. This remains at the
few-tens-of-\si{\pico\second} level throughout the glass and rises abruptly from
\SIrange{53}{351}{\pico\second} between \SI{800}{} and \SI{850}{\kelvin};
$T_g^{\mathrm{relax}}=\SI{850}{\kelvin}$ is assigned to the first plateau above the jump.
This quantity diagnoses the onset of structural relaxation on the simulation timescale rather
than measuring a relaxation time, since the drift that can be resolved depends on how long
each plateau is observed; it is reported as corroboration of the mean-based determinations in (a) and (b). Green bands mark the experimental $T_g$ as in Fig.~(\ref{fig:tg-overview}).}
    \label{fig:signatures}
\end{figure}

The two estimators are constructed independently---one from the variance of the enthalpy at
fixed temperature, the other from the temperature derivative of its mean---so their agreement
provides an internal consistency check. In particular, they return step heights of \SI{\dCpFluct}{} and
\SI{\dCpSlope}{\joule\per\mole\per\kelvin}. The residual difference is smaller than the systematic uncertainty
associated with the choice of baseline windows, which dominates the error on
$\Delta C_p$ because the supercooled branch is defined by only the two highest plateaus. The transition is corroborated by two further signatures obtained from the same simulations and shown in Fig.~(\ref{fig:signatures}): a kink in the mean potential energy at $T_g^{U}=\SI{\TgU}{\kelvin}$ (\SI{\TgUerr}{\kelvin} standard deviation over replicas) and a kink in the mean atomic volume at $T_g^{V}=\SI{\TgV}{\kelvin}$, the latter accompanied by an increase of the thermal expansivity from \SI{\AlphaGlass e-4}{} to \SI{\AlphaLiquid e-4}{\cubic\angstrom\atom^{-1}\kelvin^{-1}}. These are mean-based determinations and are therefore the most robust of the set; the calorimetric onset in Fig.~(\ref{fig:tg-overview}a), which derives from a variance, lies somewhat lower. Taken together, the estimates place the transition in the range \SIrange{\TgSlope}{\TgV}{\kelvin}.

Compared with the experimental values for this alloy, \SI{\TgExpA}{} and
\SI{\TgExpB}{\kelvin} \cite{zhang2007glass,zhou2013microstructure}, the simulated transition is higher by a factor of
\numrange{\TgRatioLo}{\TgRatioHi}. This offset is expected rather than anomalous. The glass
transition is kinetic in origin---it marks the temperature at which the structural relaxation
time becomes comparable to the observation time---and the observation windows differ by some
twelve orders of magnitude between the present simulations
($\sim\SI{e2}{\pico\second}$ per state point) and a scanning calorimetry experiment
($\sim\SI{e2}{\second}$). A shift of this magnitude is typical for metallic glass formers. By
contrast $\Delta C_p$, being thermodynamic rather than kinetic, depends far more weakly on
the observation timescale and is the more directly comparable quantity.

That the glass branch below \SI{\GlassEdge}{\kelvin} is a genuine non-equilibrium measurement deserves emphasis, since it is common to both simulation and experiment. Direct inspection of the plateau enthalpies (Supplementary Fig.~(\ref{SI-fig:htraces})) shows three regimes which directly relate to the character of $C_p(T)$ fluctuation errors: below \SI{\GlassEdge}{\kelvin} the enthalpy is stationary and the replicas fluctuate independently; at \SI{1000}{\kelvin} and above it relaxes upward and saturates within \SIrange{100}{150}{\pico\second}, so those plateaus reach equilibrium; while at \SI{850}{} and \SI{900}{\kelvin} all replicas drift downward coherently as the glass ages. Fits to saturating and non-saturating relaxation laws would distinguish these cases quantitatively, and the baselines used in $\Delta C_p$ were restricted to the temperatures that are either arrested or genuinely equilibrated. The crossover between aging and equilibration, which falls between \SI{900}{} and \SI{1000}{\kelvin}, constitutes a further and purely dynamical estimate of the transition, consistent with the thermodynamic ones. Fig.~(\ref{fig:tg-overview}b) addresses how much of this depends on the measurement protocol itself. The same samples were heated in single uninterrupted ramps at \SIlist{10;5;1.7;0.6}{\kelvin\per\pico\second}, with $C_p$ taken solely from $\mathrm{d}H/\mathrm{d}T$ along the scan, so that the heating schedule is the only difference from the plateau measurement.

In the glass, the ramps converge cleanly and monotonically towards the equilibrium result. At
\SI{300}{\kelvin} the apparent heat capacity rises from
\SIlist{\CpGlassRampA;\CpGlassRampB;\CpGlassRampC;\CpGlassRampD}{\joule\per\mole\per\kelvin}
as the rate is lowered through \SIlist{10;5;1.7;0.6}{\kelvin\per\pico\second}, against a plateau
value of \SI{\CpGlassStep}{\joule\per\mole\per\kelvin}; by \SI{500}{\kelvin} every rate has
converged. At the fastest rate the structure simply cannot absorb enthalpy quickly enough to
follow the imposed temperature, and the apparent $C_p$ falls far below the Dulong--Petit
value. The vibrational heat capacity of the glass is therefore accessible to a direct scan,
provided the rate is low enough, and the temperature at which a given rate becomes adequate
tracks the local equilibration time measured on the plateaus
(\SIrange{\EqGlassLo}{\EqGlassHi}{\pico\second} in the glass, Supplementary
Eq.~\eqref{SI-eq:si-aging}).
Through the transition, the situation is qualitatively different, and the ramps do not
converge. At \SI{900}{\kelvin} every rate examined returns
$\approx\SI{\CpNineRamp}{\joule\per\mole\per\kelvin}$ against the equilibrium
\SI{\CpNineStep}{\joule\per\mole\per\kelvin}, and even at
\SI{\RateSlow}{\kelvin\per\pico\second}---which spends \SI{\RampPerHundred}{\pico\second} per
\SI{100}{\kelvin}, longer than the \SIrange{\EqHiLo}{\EqHiHi}{\pico\second} that the plateaus
require to equilibrate---the scan reaches only
\SI{\CpKiloRamp}{\joule\per\mole\per\kelvin} at \SI{1000}{\kelvin} against
\SI{\CpKiloStep}{\joule\per\mole\per\kelvin}. The deficit reappears as an overshoot above
\SI{1100}{\kelvin}, the enthalpy not absorbed at the transition being released at higher
temperature. A criterion based on comparing the ramp's dwell time per unit temperature with
the plateau equilibration time is therefore too permissive here: the plateau value measures
how long a system held at \emph{fixed} temperature takes to relax, whereas a ramp
continuously displaces the target, and the relaxation time is itself growing steeply through
this region, so the lag accumulates rather than saturating.

The practical conclusion is that a direct continuous scan recovers the vibrational heat
capacity of the glass at modest rates, but does not recover $\Delta C_p$ or $T_g$ at any rate
accessible to this class of simulation---the stepwise protocol used here corresponds to an
effective rate of only $\approx\SI{\RateStair}{\kelvin\per\pico\second}$ while additionally
providing equilibrium sampling at every temperature. This provides a quantitative
justification for the plateau approach rather than a convention.
Finally, repeating the slowest ramp with a second interatomic potential, each acting on
glasses generated by its own quench, leaves the result essentially unchanged: \PotA{} and
\PotB{} agree to within \SI{\PotDiffTyp}{\joule\per\mole\per\kelvin} across most of the range
and everywhere within the replica scatter, the largest single-temperature difference being
\SI{\PotDiffMax}{\joule\per\mole\per\kelvin} at \SI{1100}{\kelvin} where the uncertainties are
themselves comparable. The heat capacity and the location of the transition are thus robust
to this change of description, encompassing both the energetics and the quenched structure
each potential produces.


\subsection{Performance across compositions}\label{sec:Performance across compositions}
\subsubsection{MD-quench benchmark}
\begin{figure}[t]
    \centering
    \includegraphics[width=0.99\textwidth,height=0.27\textheight,keepaspectratio]{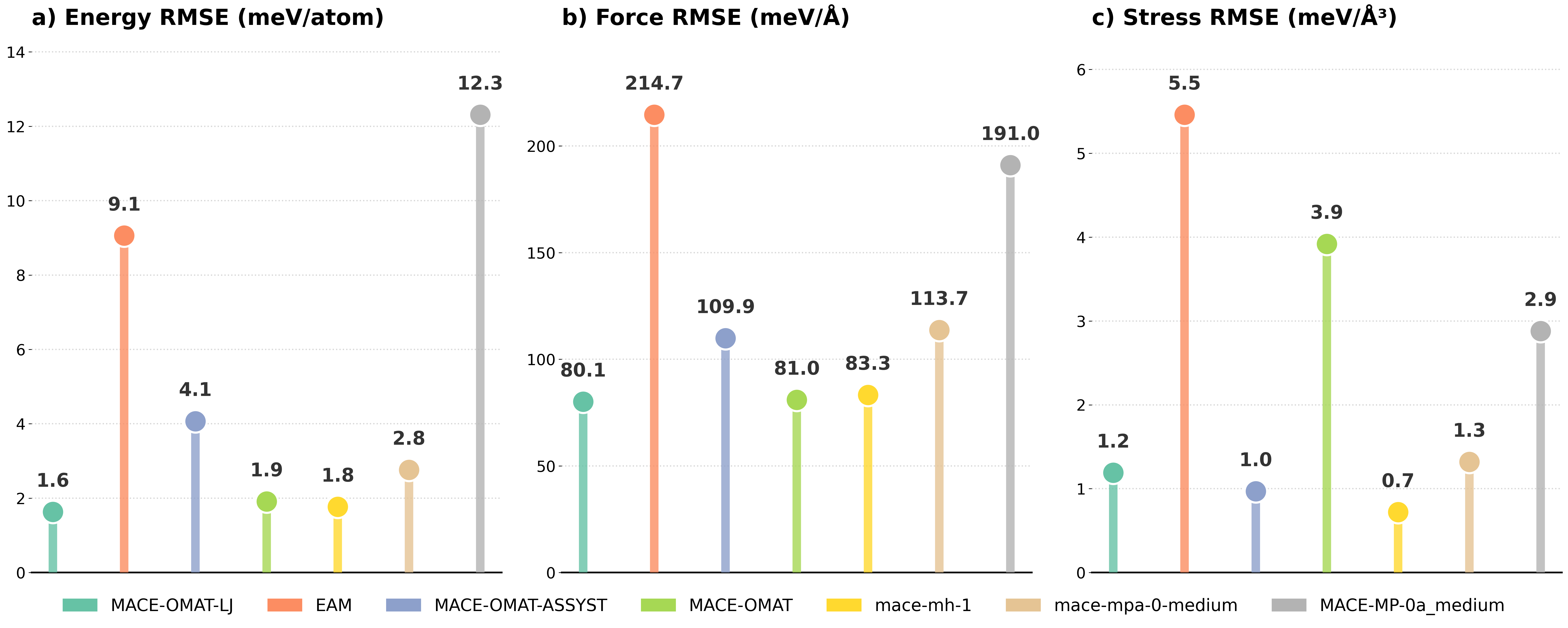}
    \caption{Global prediction error comparison of the evaluated interatomic potentials across the sampled compositional space. The lollipop charts present the cumulative Root Mean Square Error (RMSE) for (a) energy, (b) forces, and (c) stresses, calculated relative to ground-truth static DFT evaluations. These global metrics are aggregated over all representative compositions within the good glass-forming ability (GFA) region, encompassing 10 independent molecular dynamics quenching replicas per composition with 5 frames extracted per trajectory. Lower values indicate higher overall predictive accuracy and closer agreement with the DFT baseline across the analyzed phase space. Supplementary Fig.~(\ref{SI-fig:compositions_pred}) shows the explicit distribution of these errors across the Cu-Zr-Al ternary space for the focus subset of potentials.}
    \label{fig:Global_RMSE_Comparison}
\end{figure}

Moving beyond the reference stoichiometry, we evaluated the models' extrapolation capabilities across a diverse compositional space within the expected Glass Forming Ability (GFA) region.
 Starting from the quenching dynamics, Fig.~(\ref{fig:Global_RMSE_Comparison}) aggregates the global prediction errors on a series of MD-quenching frames computed across 40 different sampled compositions against DFT. 

\begin{figure}[H]
    \centering
    \includegraphics[width=0.99\textwidth]{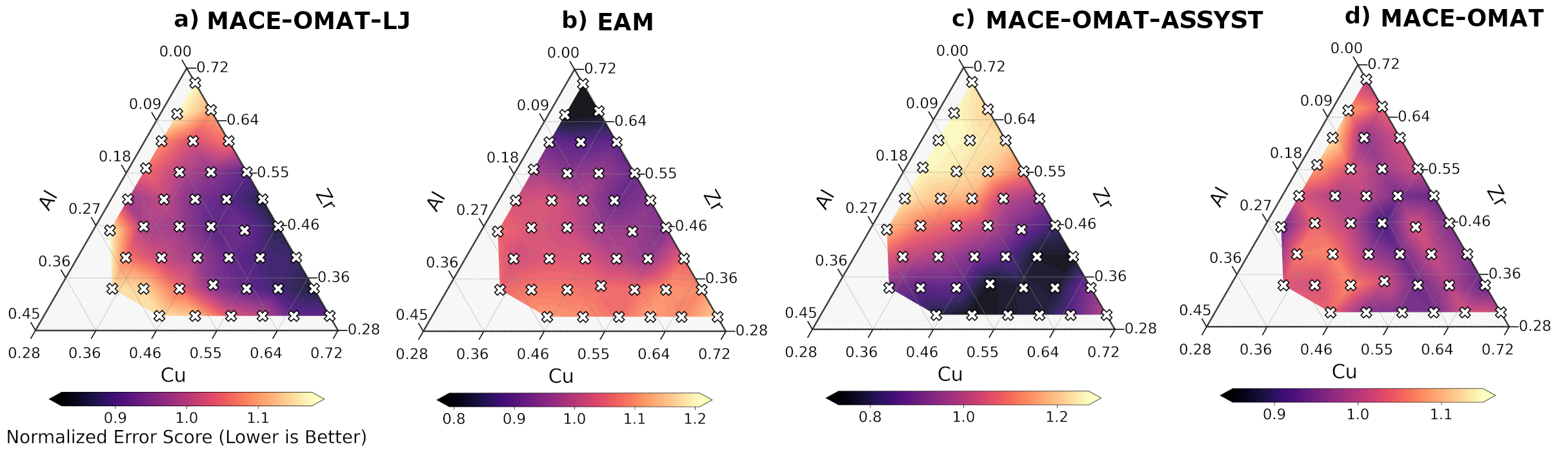}
    \caption{Normalized error score in the ternary space, computed as the average of the normalized errors over energy, force and stress: for a given composition $X$,  $\text{NES}_X=({\epsilon_{E,X} + \epsilon_{F,X} + \epsilon_{S,X}})/{3} $, where $\epsilon_{E,X}=\text{RMSE}_{E,X}/\langle \text{RMSE}_E \rangle_X$, denoting $\langle \cdot \rangle_X$ the average over all compositions. The labels a), b), c), d) have a 1:1 correspondence with the potentials in Supplementary Fig.~(\ref{SI-fig:compositions_pred}).}
    \label{fig:compositions_rank}
\end{figure}

The results highlight the exceptional out-of-the-box generalization power of the foundational architecture. Focusing first on our subset of interest, the pre-trained MACE-OMAT and its LJ-specialized variant (MACE-OMAT-LJ) maintain remarkable accuracy across the varied stoichiometries, achieving cumulative energy RMSEs of $1.9$ and $1.6$ meV/atom, respectively, and force RMSEs of $81.0$ and $80.1$ meV/Å. In stark contrast, the classical EAM potential fails to generalize efficiently, yielding a massive global energy error of $9.1$ meV/atom and a force RMSE of $214.7$ meV/Å. Notably, fine-tuning the foundation architecture on the automated small-cell database (MACE-OMAT-ASSYST) degrades its intrinsic compositional generalization, pushing the global energy RMSE up to $4.1$ meV/atom and the force RMSE to $109.9$ meV/Å. 
The explicit distribution of these errors across the Cu-Zr-Al ternary space is detailed in Supplementary Fig.~(\ref{SI-fig:compositions_pred}), while Fig.~(\ref{fig:compositions_rank}) presents a normalized error score (NES) useful to grasp where each potential from the focus set, separately, is working best in the ternary good GFA space. 
While MACE-OMAT and MACE-OMAT-LJ provide a uniformly stable, low-error predictive surface across all evaluated concentrations, the EAM potential suffers from high errors throughout the entire mapped phase space. Furthermore, a closer look at the NES reveals that the ASSYST-fine-tuned model's predictive accuracy degrades noticeably as concentrations shift toward the boundaries of the ternary space, indicating that fine-tuning on small periodic cells disrupts the broad compositional coverage originally present in the pre-trained backbone. 

Finally, to validate dynamic stability,
we tracked the thermodynamic evolution during MD quenching for specific target compositions and computed their global normalized error scores, which we report in Supplementary Fig.~(\ref{SI-fig:compositions_trajs}). For the MACE-OMAT-LJ potential, the thermodynamic trajectories demonstrate reliability with respect to DFT; even for the ``Worst'' performing composition, the model exhibits minimal deviation from the ground-truth (represented by small vertical error bars) and tightly bound configurational variance across 10 independent replicas (narrow shaded regions) over the entire cooling range. The EAM potential instead exhibits larger absolute energy offsets and force evaluation errors across all ranked compositions.

Going back to the comparison between different out-of-the-box foundation architectures, Fig.~(\ref{fig:Global_RMSE_Comparison}) strikingly highlights the critical role of the underlying pre-training database. As discussed, models pre-trained on the OMAT database, which explicitly includes high-temperature liquids and amorphous configurations—namely, MACE-OMAT and mace-mh-1—exhibit exceptional zero-shot baseline accuracy. 
Conversely, MACE-MP-0a\_medium, which is pre-trained predominantly on near-equilibrium crystalline structures from the MPTrj database, fails to generalize to structural disorder. It yields the highest global energy error of the entire set at $12.3$ meV/atom and massive force deviations reaching $191.0$ meV/Å. While adding the sAlex dataset to the pre-training mix (mace-mpa-0-medium) noticeably recovers performance—dropping energy errors to $2.8$ meV/atom and forces to $113.7$ meV/Å—it still fundamentally lags behind the architectures exposed to true disorder during pre-training.
\subsubsection{Elastic property - discovery task assessment}
Having established the accuracy of the potential on its training targets, we now test how well it \emph{generalizes} to a bulk mechanical property that was not fitted directly: the isotropic elastic response of the as-quenched glass, evaluated across the selected Cu--Zr--Al composition space. 
\begin{figure}[t]
    \centering
    \includegraphics[width=0.80\textwidth]{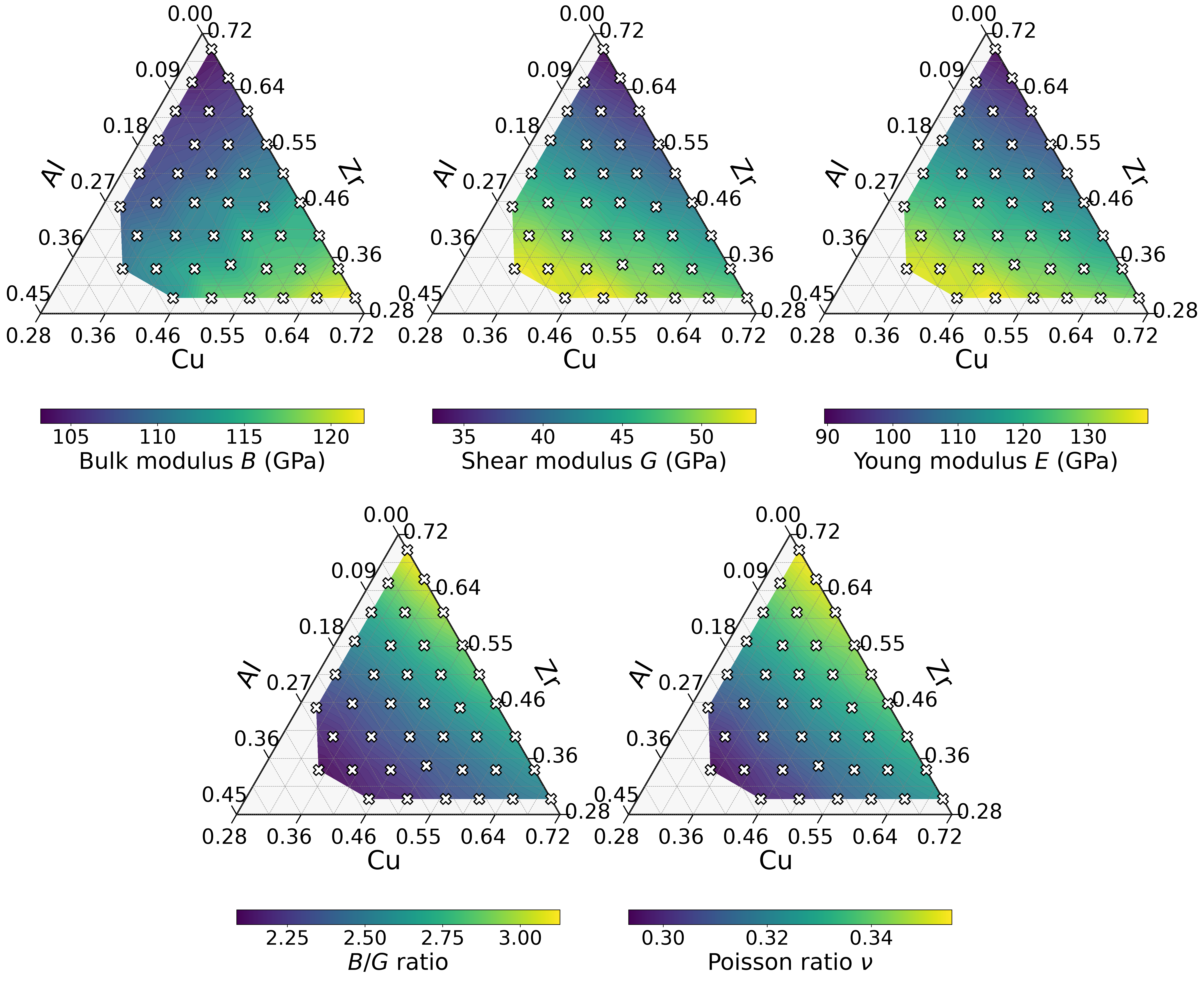}
    \caption{Composition dependence of the isotropic elastic moduli of the Cu--Zr--Al metallic glasses predicted by the MACE-OMAT-LJ model. Each panel maps one property over the sampled region of the ternary (bottom/right/left axes: Cu/Zr/Al mole fractions): bulk modulus~$B$, shear modulus~$G$, Young modulus~$E$, the $B/G$ ratio (Pugh ductility indicator) and Poisson ratio~$\nu$. For every configuration, the full $6\times6$ elastic tensor is obtained from finite deformations (normal strain up to \SI{1}{\percent}, shear strain up to \SI{6}{\percent}, five deformations per mode) at the as-quenched geometry, and reduced to the isotropic moduli through the Voigt--Reuss--Hill average. The color field is a linear (barycentric) interpolation of the replica-averaged prediction over the \num{40} sampled compositions (white crosses), each averaged over \num{10} independent quench replicas. $B$, $G$ and $E$ are in \si{\giga\pascal}; $B/G$ and $\nu$ are dimensionless. Supplementary Fig.~(\ref{SI-fig:elastic_ternary_uncertainty}) maps the standard deviation of each modulus over the ten quench replicas.}
    \label{fig:elastic_ternary_mean}
\end{figure}
We sample \num{40} compositions spanning the ternary (Cu~\SIrange{30}{65}{\percent}, Zr~\SIrange{30}{70}{\percent}, Al~\SIrange{0}{30}{\percent}), each represented by \num{10} statistically independent quench replicas. For every configuration, the full $6\times6$ elastic stiffness tensor is computed from finite deformations at the as-quenched geometry and reduced to the isotropic bulk~($B$), shear~($G$) and Young~($E$) moduli, the Poisson ratio~($\nu$) and the $B/G$ Pugh ductility indicator through the Voigt--Reuss--Hill average. Averaging over the replica ensemble yields a smooth composition map of each modulus (Fig.~(\ref{fig:elastic_ternary_mean})); the replica scatter quantifies the sensitivity to the underlying structural disorder (Supplementary Fig.~(\ref{SI-fig:elastic_ternary_uncertainty})); and a direct comparison with ultrasonic measurements collected from the literature tests the quantitative accuracy (Fig.~(\ref{fig:ternary-exp-zscore})).

The predicted moduli vary smoothly and monotonically over the sampled region (Fig.~(\ref{fig:elastic_ternary_mean})), spanning $B \approx \SIrange{103}{121}{\giga\pascal}$, $G \approx \SIrange{33}{53}{\giga\pascal}$, $E \approx \SIrange{90}{139}{\giga\pascal}$, $\nu \approx \numrange{0.29}{0.36}$ and $B/G \approx \numrange{2.1}{3.1}$, values characteristic of Cu--Zr-based bulk metallic glasses. Two physical trends emerge. First, enriching the alloy in Cu stiffens the glass, raising both $B$ (from $\approx\SI{103}{\giga\pascal}$ at \SI{30}{\percent}~Cu to $\approx\SI{122}{\giga\pascal}$ at \SI{65}{\percent}~Cu) and $G$. Second, Al additions act predominantly on the shear response: along the (Cu$_{50}$Zr$_{50}$)$_{100-x}$Al$_x$ pseudo-binary, adding \SI{10}{\percent}~Al leaves $B$ essentially unchanged ($\approx\SI{111}{}\rightarrow\SI{112}{\giga\pascal}$) while raising $G$ from $\approx\SI{39}{}$ to $\approx\SI{44}{\giga\pascal}$ and $E$ from $\approx\SI{105}{}$ to $\approx\SI{117}{\giga\pascal}$, and lowering $B/G$ (\num{2.85}$\rightarrow$\num{2.53}) and $\nu$ (\num{0.343}$\rightarrow$\num{0.326}). The Al-induced stiffening of $G$ and $E$ reproduces the trend reported experimentally for the same alloys~\cite{yu2008poisson,li2009effect}.

The statistical robustness of these predictions is documented in Supplementary Fig.~(\ref{SI-fig:elastic_ternary_uncertainty}), which maps the standard deviation of each modulus over the ten quench replicas. The spreads are small throughout the ternary---$\sigma_B,\sigma_G \lesssim \SI{3}{\giga\pascal}$, $\sigma_E \lesssim \SI{4}{\giga\pascal}$ (a few percent of the mean) and $\sigma_\nu \lesssim \num{0.004}$---and are everywhere much smaller than the composition-driven variation of Fig.~(\ref{fig:elastic_ternary_mean}). The elastic response is therefore essentially insensitive to the particular realization of the quenched glass, so that composition, rather than quench-to-quench disorder, controls the moduli; these replica statistics are what propagate into the $\pm1\sigma$ error bars of the validation below.

Finally, Fig.~(\ref{fig:ternary-exp-zscore}) confronts the predictions with experiment at every measured composition, mapping the standardized residual $z=(\overline{\mathrm{model}}-\mathrm{exp})/\sigma_{\mathrm{model}}$---with $\overline{\mathrm{model}}$ and $\sigma_{\mathrm{model}}$ the replica mean and standard deviation of the prediction at that composition---directly onto the ternary; the model value is again obtained by interpolation of the \num{40}-composition grid, reducing to the direct prediction on a sampled node. 
\begin{figure}[t]
  \centering
  \includegraphics[width=\textwidth]{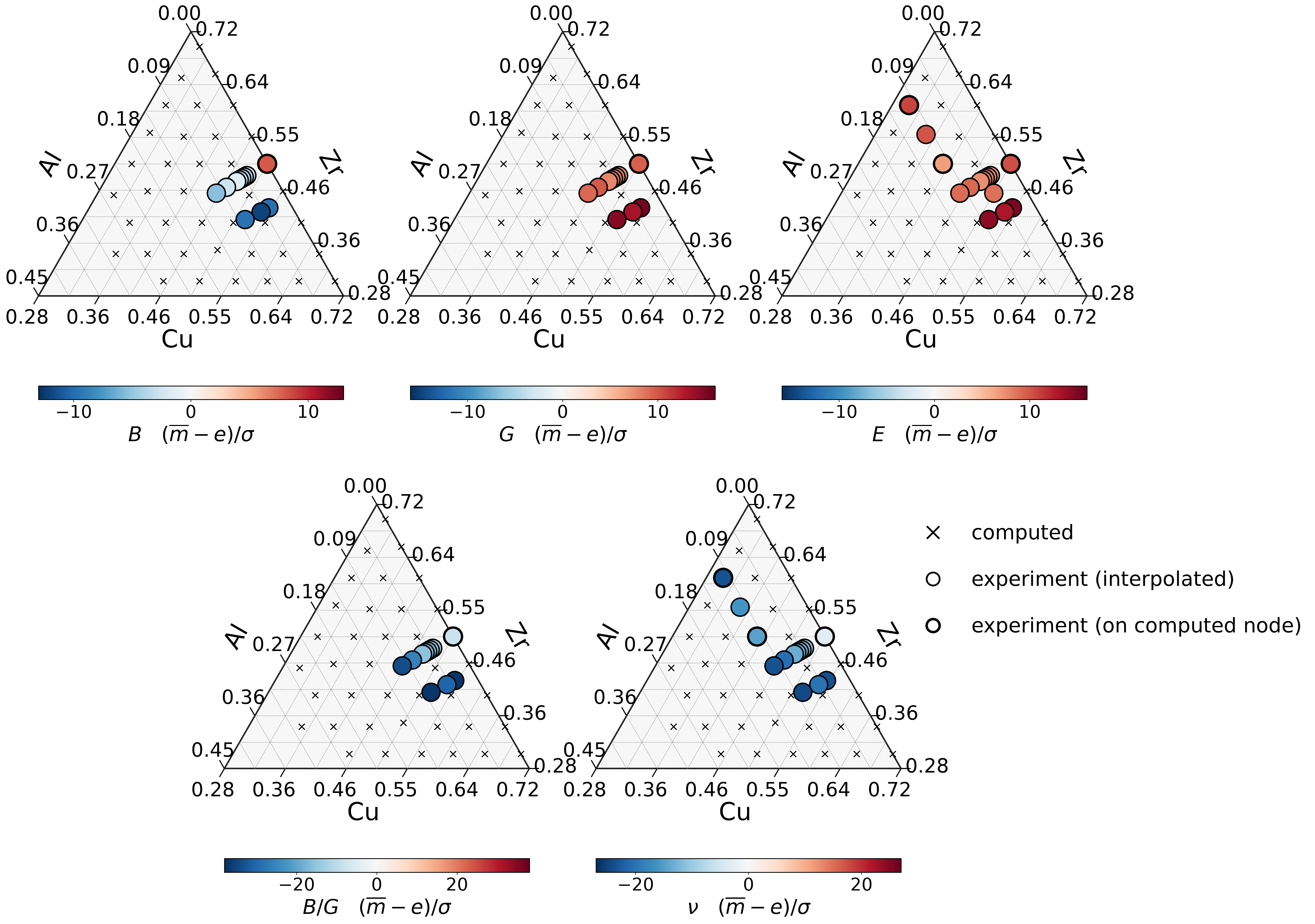}
  \caption{%
    Spatially resolved validation of the MACE elastic predictions against experiment across the Cu--Zr--Al ternary, expressed as the \emph{standardized residual}. Each panel maps one property (bulk modulus~$B$, shear modulus~$G$, Young modulus~$E$, the $B/G$ Pugh ratio and Poisson ratio~$\nu$) on the same zoomed simplex and axis convention as Fig.~(\ref{fig:elastic_ternary_mean}) and Supplementary Fig.~(\ref{SI-fig:elastic_ternary_uncertainty}) (bottom/right/left $=$ Cu/Zr/Al mole fraction). Black crosses mark the \num{40} computationally sampled compositions; filled discs mark the experimentally measured compositions, colored by $z = (\overline{\mathrm{model}}-\mathrm{exp})/\sigma_{\mathrm{model}}$, where $\overline{\mathrm{model}}$ and $\sigma_{\mathrm{model}}$ are the replica mean and replica standard deviation of the prediction at that composition. The model value is obtained by the same linear (barycentric) interpolation of the computed grid used for the property maps, reducing to the direct prediction when the composition coincides with a sampled node (thick disc edge) rather than being interpolated (thin disc edge). The diverging scale is symmetric about zero and set independently for each panel: red marks compositions where the model over-predicts, blue where it under-predicts, and near-white agreement to within the replica scatter. Since $\sigma_{\mathrm{model}}$ is only a few \si{\giga\pascal}, $|z|\gtrsim 1$ over almost the whole measured region, i.e.\ the deviations are statistically significant and systematic rather than a reflection of quench-to-quench scatter. A property is shown only at the compositions for which it was reported, and regions of the ternary with no nearby measurement are left blank. The experimental set comprises the (Cu$_{50}$Zr$_{50}$)$_{100-x}$Al$_x$ series ($x=0$--$10$) \cite{fan2006thermophysical,yu2008poisson}, the (Cu$_{56}$Zr$_{44}$)$_{100-x}$Al$_x$ series \cite{li2009effect}, a Zr--Cu--Al series at fixed \SI{10} {\percent}~Al with varying Zr/Cu ratio \cite{idriss2015evolution}, and Cu$_{50}$Zr$_{45}$Al$_5$ \cite{plummer2009elastic}. The maps localize the shear over-stiffness of the potential: $G$ and $E$ are systematically over-predicted (deep red), $B/G$ and $\nu$ correspondingly under-predicted (deep blue), whereas the bulk modulus~$B$ is reproduced within a few standard deviations (mixed, near white to blue).%
  }
  \label{fig:ternary-exp-zscore}
\end{figure}
Measuring the discrepancy in units of the quench-replica spread is what makes this view diagnostic: since $\sigma_{\mathrm{model}}$ amounts to only a few \si{\giga\pascal} (Supplementary Fig.~(\ref{SI-fig:elastic_ternary_uncertainty})), $|z|$ reaches values of order ten across the measured region, so the gap cannot be ascribed to quench-to-quench scatter. The panels separate cleanly by sign. The shear-dominated moduli are over-predicted at every single composition, $z\approx+7$ to $+15$ for both $G$ and $E$ (bias $+\SI{10}{\giga\pascal}$, \SI{31}{\percent}, and $+\SI{23}{\giga\pascal}$, \SI{27}{\percent}, respectively), and the two ratios built on them follow mechanically: $B/G$ and $\nu$ are under-predicted everywhere ($\approx\num{-1.0}$ and $\Delta\nu\approx\num{-0.04}$, \SI{10}{\percent}), and carry the largest excursions of the whole figure, down to $z\approx\num{-37}$ and $z\approx\num{-25}$, simply because their replica spreads are the tightest ($\sigma_\nu\lesssim\num{0.004}$). The bulk modulus stands apart: it stays pale over most of the simplex at $|z|\approx2$--$5$, corresponding to a mean absolute error of \SI{6.6}{\giga\pascal} (\SI{6}{\percent}) and a mild underestimate, and turns red only at the Al-free Cu$_{50}$Zr$_{50}$ binary, the one composition where $B$ is over-predicted. Resolving the residuals spatially also localizes where the description degrades most: the Cu-rich corner sampled by the (Cu$_{56}$Zr$_{44}$)$_{100-x}$Al$_x$ series~\cite{li2009effect} carries the larger error in the $B$, $G$, $E$ and $B/G$ panels ($z\approx\num{-10}$ to $\num{-12}$ for $B$, $+13$ to $+15$ for $G$ and $E$), while the Al-containing (Cu$_{50}$Zr$_{50}$)$_{100-x}$Al$_x$ pseudo-binary remains the best described. Crucially, within each family the model tracks the experimental ordering and composition trend while sitting on a nearly constant offset (for completeness see Supplementary Fig.~(\ref{SI-fig:elastic_validation_vs_exp})), so what the maps expose is a single, spatially smooth shear-stiffness bias rather than a complete breakdown of transferability. Partially, this offset is expected for multiple reasons, including small sample size and fast quenching. 
\clearpage
\section{Discussion}\label{sec:Discussion}
Our results indicate that pre-training coverage is a key factor in whether a machine-learned potential can describe structural disorder---at least for the architecture examined here---more so than the amount of system-specific data it is shown afterwards.
The pre-trained MACE-OMAT model, which has not been exposed to BMG configurations, follows the quench of the target glass to within $1$--$2$~meV/atom in energy and $\approx 80$~meV/\AA{} in force (Fig.~(\ref{fig:trajectory_Target_analysis}a)), generalizing such performance level to the entire good GFA region in the ternary space, reproduces reasonable partial radial distribution functions (Fig.~(\ref{fig:trajectory_Target_analysis}b)) of the target glass, predicts the elastic moduli at near-DFT fidelity (Fig.~(\ref{fig:elastic_nes_grid})), gives a mechanical response indistinguishable from that of its BMG-fine-tuned variant up to and beyond yielding (Fig.~(\ref{fig:shear_combined})), and captures the signature of the glass transition in the heat capacity. It is worth underlining that, for this model, none of this required a dedicated training campaign and its pre-trained stage was not exposed to quenched glassy metals to the best of our knowledge.

The comparison across pre-training corpora makes the argument sharp. The same architecture pre-trained on the near-equilibrium MPTrj data\cite{Deng2023} fails outright ($12.3$~meV/atom, $191.0$~meV/\AA), and adding sAlex\cite{schmidt2023machine} recovers only part of the gap 
(Fig.~(\ref{fig:Global_RMSE_Comparison})) in the MD-quenching test over the ternary space. Since MPTrj$+$sAlex is the larger of the two near-equilibrium corpora and still lags, the relevant quantity is not dataset size but explicit coverage of liquid and off-equilibrium environments: physically, the model must have learned how to map the short- to medium-range order of the local chemical environment of a glass, leveraging these mixed and rich representations.

We present supporting evidence of the available EAM baseline failing in dynamical tests. The EAM potential\cite{EAM} (Sec.~\ref{sec:EAM focus}) was fitted to a few hundred DFT configurations of $96$--$128$ atoms at a single composition, below the size argued to be the minimum for bulk metallic glasses, 
and is by construction non-exploratory. Its total $g(r)$ looks acceptable while its partials slightly misplace the chemical short-range order, overestimating the first Cu--Al and Zr--Zr peaks and underestimating Al--Al and Cu--Cu. Its elastic description produces a glass too stiff in shear and too compliant hydrostatically ($E$ and $G$ overestimated, $B$ and $\nu$ underestimated), giving an aggregate score of $\mathrm{NES}=2.54$ against $0.40$--$0.53$ for the MACE family. That error grows systematically from slowly to rapidly quenched samples, whereas the OMAT-based foundation models are almost rate-insensitive, indicating that the embedded-atom description captures well-relaxed inherent structures but degrades on the higher-energy configurations retained at fast quench. Finally, as MD-quenching tests show, it does not transfer in composition ($9.1$~meV/atom, $214.7$~meV/\AA), which is what disqualifies it for screening applications.

Fine-tuning on target-composition disorder improves, but unevenly and by little. Specializing the backbone on the LJ database moves the energy and force errors from $1.9$ to $1.6$~meV/atom and from $81.0$ to $80.1$~meV/\AA{}---changes of a few percent. The genuine gains are elsewhere: the stress error drops from $3.9$ to $1.2$~meV/\AA$^3$ and the elastic score improves from $0.53$ to $0.40$, the best of the set. The natural reading is that the pre-trained model is already interpolating these environments, so that further target data enters a regime of diminishing returns. For this system the bottleneck is no longer training data.

Specializing instead on automatically generated small cells is found to be harmful. The underlying question of whether a very rich exploration of (rattled) stoichiometries and (sheared) symmetries in the ternary CuZrAl space could mimic the very local environment of real glasses has a negative answer: the ASSYST-fine-tuned model degrades the MD-quench generalization test to $4.1$~meV/atom and $109.9$~meV/\AA{}  while leaving the elastic score untouched ($0.531$ against $0.533$). The degradation is structured: it grows towards the boundaries of the ternary (Fig.~(\ref{fig:compositions_rank})), eroding the balanced compositional coverage acquired during pre-training. The ASSYST and LJ distributions \emph{do} overlap in the pre-trained model's own embedding space, but coverage in a learned representation is found to be a necessary but not a sufficient condition for successful specialization, just as are good energy, force and stress metrics; deployment in long MD-production better exposes improvements or degradation against emerging dynamics-related phenomena.   

Structure and the hydrostatic response are consistent: all MACE variants superpose on every atomic pair, and the bulk modulus is reproduced to $6.6$~GPa on average ($6\%$). Shear-dominated quantities carry a systematic offset, with $G$ and $E$ overestimated by $31\%$ and $27\%$ and the derived $B/G$ and $\nu$ correspondingly low (Fig.~(\ref{fig:ternary-exp-zscore})). That offset is spatially smooth across the ternary and largest in the Cu-rich corner, which identifies it as a single correctable bias rather than a breakdown of transferability. Part of it is not attributable to the potential: DFT is itself offset from experiment for these observables, setting a floor on the achievable agreement, and the moduli are evaluated at the as-quenched geometry of rather small, rapidly cooled samples. The same distinction applies to the glass transition, where the simulated $T_g$ exceeds the calorimetric value by a factor $1.2$--$1.3$ (Figs.~(\ref{fig:tg-overview}) and~(\ref{fig:signatures})); this is the expected consequence of an observation window some twelve orders of magnitude shorter than a scanning experiment, and it is $\Delta C_p$, being thermodynamic rather than kinetic, that remains the directly comparable quantity.
This study underlines the capability of existing foundational architectures to capture the glass transition, both in the rigorous regime of plateau enthalpy acquisition and in very slow direct ramping. 

Sampling the glass-forming region yields smooth maps of $B$, $G$, $E$, $\nu$ and $B/G$ (Fig.~(\ref{fig:elastic_ternary_mean})) that reproduce the experimentally reported trends---Cu enrichment stiffens the glass, while Al acts selectively on the shear response, raising $G$ and $E$ at nearly constant $B$---with quench-to-quench scatter far smaller than the composition-driven variation. Composition, not the particular glass realization, controls the moduli, which is the condition that makes screening meaningful. 

These conclusions are drawn from one alloy family and one architecture family, at the PBE level, with cooling rates and observation windows far from experiment. Within those limits, we expect the implications to hold more broadly, although confirming this for other glass families---including non-metallic ones---will require further work. Foundation datasets should include strong off-equilibrium, high-T configurations by design, since the MPTrj--OMAT contrast shows this is what separates a usable model from an unusable one towards structural disorder. Because additional target data buys so little, the next substantial gain is more likely to come from architectures adapted to the non-local, medium-range character of disorder than from further fine-tuning. 





\section{Methods}\label{sec:methods}

All DFT calculations were performed using the Vienna Ab-initio Simulation Package (VASP) version 6.5.1~\cite{Kresse1996, Kresse1999} without spin polarization and employing the Perdew–Burke–Ernzerhof (PBE) exchange-correlation functional~\cite{Perdew1996}.
The systems under study were composed of different concentrations of aluminum, copper, and zirconium, for which the suggested VASP pseudopotentials were employed. 
Details on the KPOINTS, ENCUT and smearing used in the calculations reported in this manuscript are shown in Supplementary Tab.~\ref{SI-tab:assyst_parameters}.

Every molecular dynamics simulation was based on the use of LAMMPS version 10 Sep 2025~\cite{LAMMPS}, and the simulations with the MACE models utilized the ML-IAP package.


\subsection{Data generation strategies}\label{sec:Data generation strategies}

\subsubsection{The LJ-supported method}\label{sec:The LJ-supported method}
In their recent work, A. Wadowski \textit{et al.}\cite{wadowski2026efficient} deploy a framework for developing MLIPs specifically tailored for the complex, disordered structures of metallic glasses.
The process begins by utilizing a computationally inexpensive LJ surrogate model to explore the PES of the material. Accelerated sampling is achieved through swap-Monte Carlo moves, which allow the system to reach deeply supercooled and ultrastable glassy states that are typically inaccessible via standard molecular dynamics. Single-point DFT corrections are applied to the configurations generated by the surrogate model, refining energies and forces, limiting the need for exhaustive and expensive full structural relaxations. The training database is strategically built to cover a wide configurational space, including globally disordered states, various crystalline phases, and configurations subjected to volumetric expansions, compressions, and distortions. Finally, a machine-learning potential can be trained on the resulting configurations and further refined through active learning. 

In this work, we consider the original database from the publication of Wadowski \textit{et al.} \cite{wadowski2026efficient} for fine-tuning the foundation potentials, which is focused on the composition $\text{Cu}_{0.46}\text{Zr}_{0.46}\text{Al}_{0.08}$. We address this database with the suffix ``LJ'', and we keep unaltered train and test separation. 
\subsubsection{ASSYST}\label{sec:ASSYST}
The ASSYST (Automated Small SYmmetric Structure Training) \cite{poul2025automated} method is an automated framework designed to create diverse datasets for training MLIPs. The core idea is to explore the PES by generating a wide variety of crystal configurations without the need for prior specialized knowledge of the material. Training on small cells (i.e., 12 atoms) leverages the cubic scaling of DFT, providing more useful data points per unit of computing time compared to larger supercells. The authors show that a moment tensor potential (MTP) trained on small, varied cells can extrapolate remarkably well on the energetics of unseen larger structures, including point defects, surfaces, and grain boundaries. 

\begin{figure}[htbp]
    \centering
    
    \begin{subfigure}[b]{0.95\textwidth}
        \textbf{a)} 
        \phantomcaption 
        \label{fig:top} 
        \vspace{-2ex} 
        
        \centering
        \includegraphics[width=0.95\linewidth]{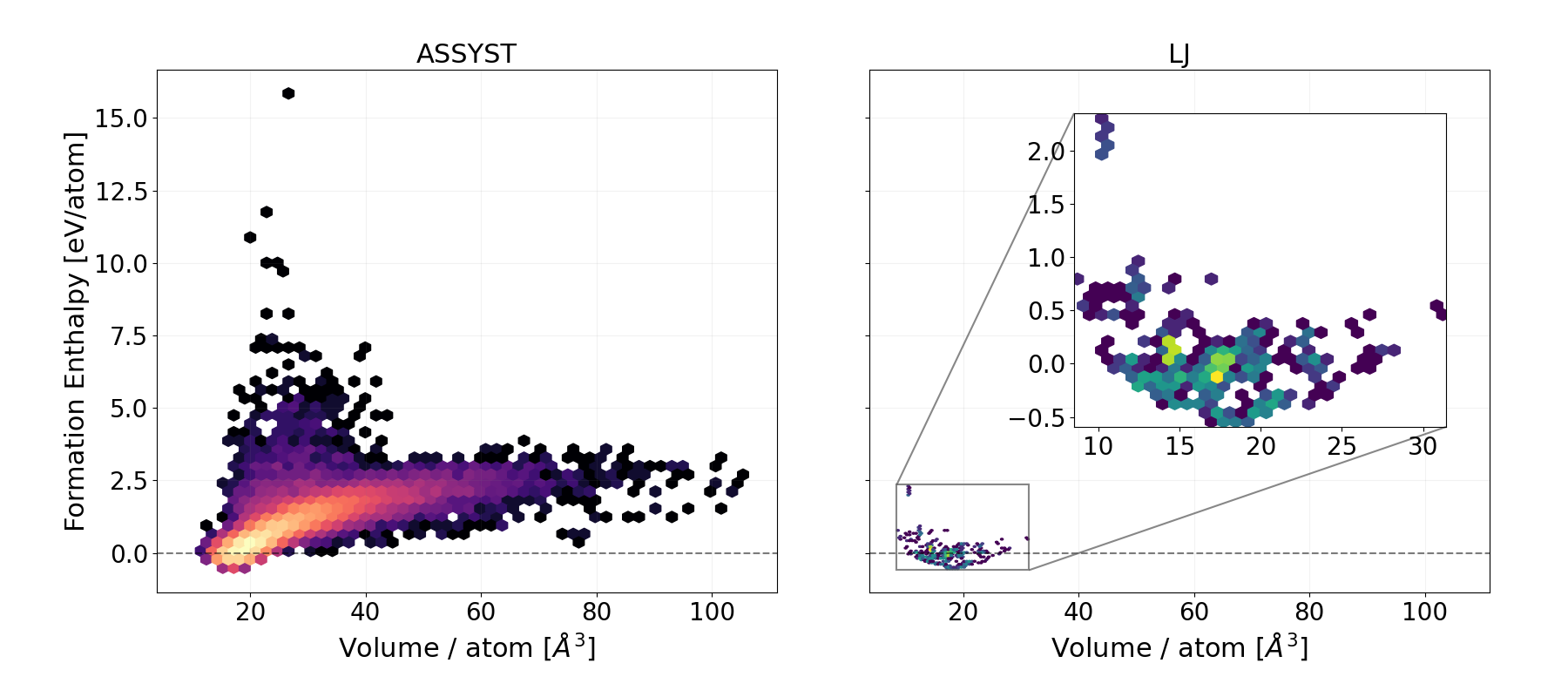}
    \end{subfigure}
    
    \vspace{-2ex} 
    
    \begin{subfigure}[b]{0.95\textwidth}
        \textbf{b)} 
        \phantomcaption 
        \label{fig:bottom}
        \vspace{-2ex} 
        
        \centering
        \includegraphics[width=0.95\linewidth]{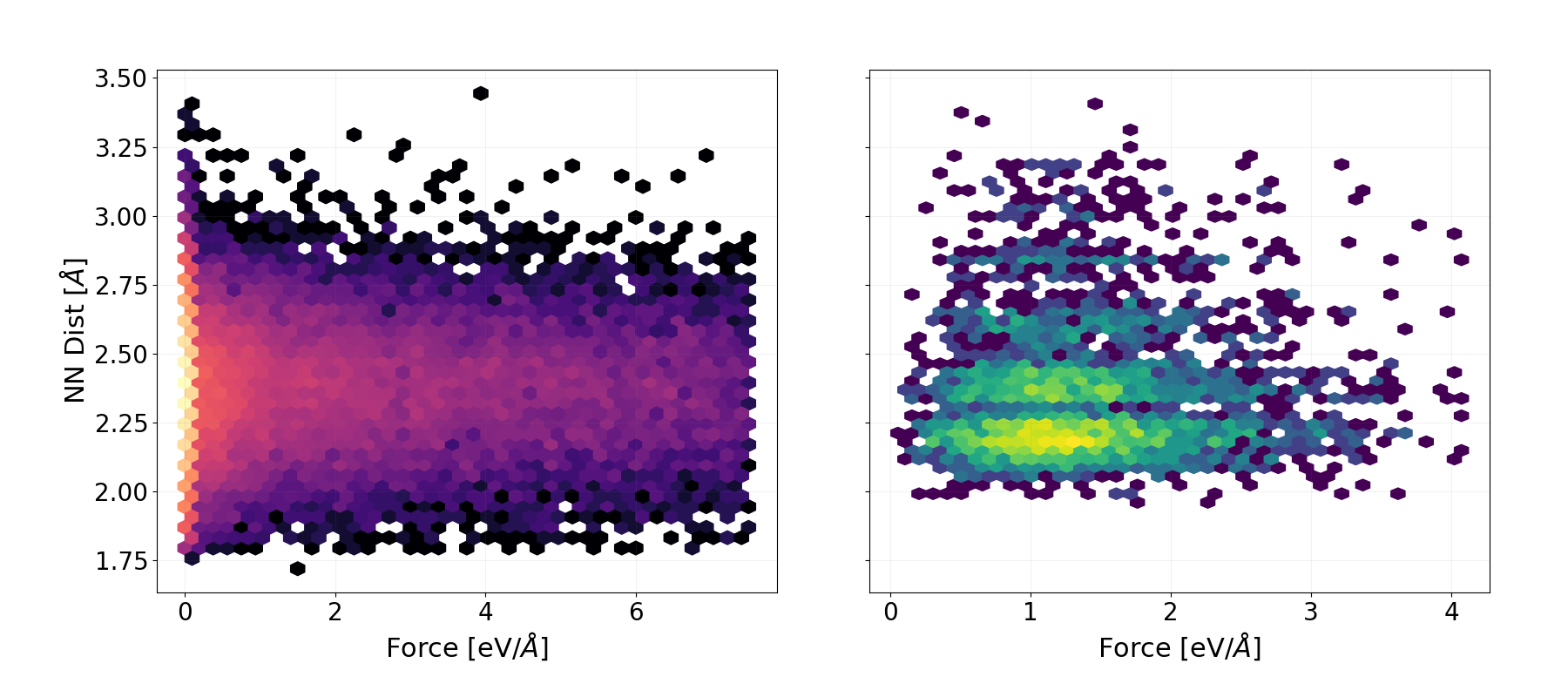}
    \end{subfigure}
    
    \caption{Comparison of structural and thermodynamic property distributions between the ASSYST and LJ databases. a) Distribution of formation enthalpy per atom as a function of volume per atom for the ASSYST dataset (left) and the LJ dataset (right), with the latter featuring an inset detailing the low-volume, low-enthalpy regime where data is found. b) Nearest-neighbor distance (NN Dist) plotted against atomic force for ASSYST (left) and LJ (right). Data density is indicated by color heatmaps, revealing the distinct structural spaces, force ranges, and phase stability bounds sampled by each database.}
    \label{fig:AS_LJ_energies_AS_LJ_forces}
\end{figure}

For a comprehensive presentation of the approach, we address the reader to the reference work from M. Poul \textit{et al.}\cite{poul2025automated}. The key workflow follows these steps (parameters in Supplementary Tab.(\ref{SI-tab:assyst_parameters})):\begin{enumerate}
    \item \textbf{Initial Generation}: The system creates random periodic structures across various space group symmetries. 
    \item \textbf{Two-Stage Relaxation}: These random structures undergo preliminary volume relaxation followed by full structural relaxation using low-convergence DFT settings.
    \item \textbf{Enhanced Sampling}: To capture the regions surrounding energy minima, the relaxed structures are modified with random atomic displacements and cell deformations.
    \item \textbf{Final Data Acquisition}: High-precision static DFT calculations are performed on all gathered configurations to produce the final energy, force, and stress data.
\end{enumerate}
The parameters used during acquisition of the ASSYST database are reported in Supplementary Tab.(\ref{SI-tab:assyst_parameters}). 
As reported in Fig.~(\ref{fig:AS_LJ_energies_AS_LJ_forces}a), the database generated via the ASSYST framework exhibits a broad, dense cloud mapping out formation enthalpies from the ground-state up to highly unstable configurations ($> 5.0\text{ eV/atom}$), stretching across volumes spanning from approximately $15$ to over $100\text{ \AA}^3/\text{atom}$. This extensive coverage reflects the design principle of ASSYST: by systematically scanning random crystal structures across multiple space groups, the approach successfully probes not only low-energy crystalline ground states, but also captures the broad potential energy landscape (PES) corresponding to highly non-equilibrium phases.
Conversely, the LJ dataset features a severely localized footprint. As zoomed in the inset, the sampled configurations are exclusively tightly bound to a narrow low-volume region ($10$ to $30\text{ \AA}^3/\text{atom}$) and a restricted low-enthalpy regime ($\Delta H_{\text{form}} \lesssim 1.0\text{ eV/atom}$), omitting highly expanded or highly unstable states, in agreement with liquid/glass thermal sampling which focuses on regions accessible under close-to-equilibrium thermodynamic paths.
Instead, Fig.~(\ref{fig:AS_LJ_energies_AS_LJ_forces}b) compares the explored forces in the two approaches. ASSYST displays a highly dense and continuous distribution of atomic forces reaching $7.5\text{ eV/\AA}$ over a wide area of nearest-neighbor (NN) distances ($1.75\text{ \AA}$ to $3.5\text{ \AA}$), a consequence of the recursive nested deformation workflow, forcing atoms closer together into highly non-equilibrium, repulsive states.
The physics-guided liquid/glass thermal baseline is restricted to a more compact region. Forces truncate around $4\text{ eV/\AA}$, and the structural environments are bounded within a narrower range of stable interatomic bonds ($2.0\text{ \AA}$ to $2.75\text{ \AA}$), as thermal paths naturally avoid high-energy barriers or unphysical atom overlaps, so they do not explore very harsh coordinate displacements. 

 Finally, Supplementary Figs.~(\ref{SI-fig:04_ternary_composition}) and (\ref{SI-fig:07_space_groups}) compare the ternary composition space and space groups coverage in the two approaches. 

\subsection{Foundation models}\label{sec:Foundation models}

\subsubsection{Screening strategy}\label{sec:Screening strategy}
A variety of foundation models are available as open source, based on different architectures and training datasets. In this work, we focus on models from the MACE\cite{batatia2022mace} family. 

\begin{figure}[t]
    \centering
    \includegraphics[width=0.99\textwidth]{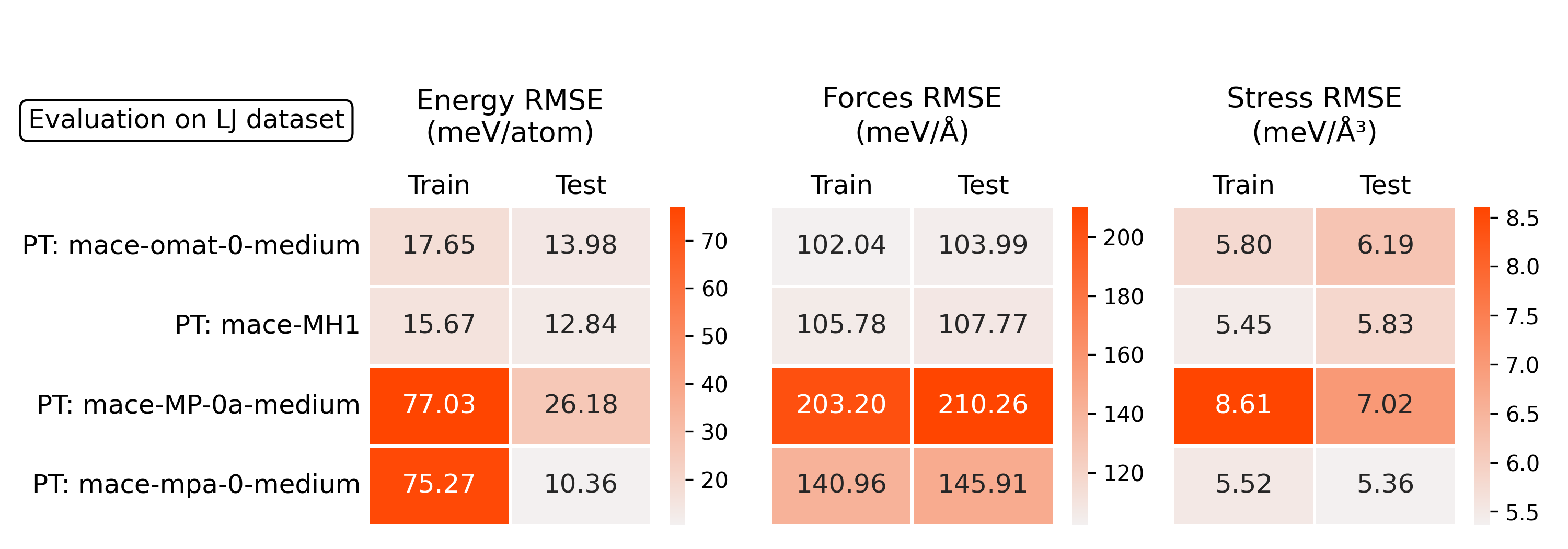}
    \caption{Evaluation of four pre-trained MACE foundation models on the LJ-test database of CuZrAl bulk metallic glasses. The heatmaps compare the Root Mean Square Error (RMSE) for energy (meV/atom), forces (meV/Å), and stress (meV/Å³) across training and test sets.}
    \label{fig:Performance_MACE_LJ_RMSE_AllPretrained}
\end{figure}

As a first strategy to rank different pre-trained (PT) models, we evaluate them on the original train and test split of the LJ database as its reference work\cite{wadowski2026efficient}, focusing on the RMSE metrics on energy, forces and cell stress. The results in Fig.~(\ref{fig:Performance_MACE_LJ_RMSE_AllPretrained}) show the mace-omat and mace-MH1 models being more competitive. In the context of this work, the reason is to be found in the larger pre-training database of the two models with respect to the others. As shown in Supplementary Tab.~(\ref{SI-tab:macemodelspt}), these models have been trained on the OMAT\cite{barroso2024open} database, which spans $\sim$100M structures, including off-equilibrium, high-temperature configurations, critical for learning the PES in liquids and amorphous phases. Such structures clearly represent an advantage when it comes to predicting the properties of amorphous metals, with respect to simpler and less populated databases mainly focusing on near-equilibrium structures like the MPTrj\cite{Deng2023} or the sAlex\cite{schmidt2023machine} databases.

Foundational models and their zero-shot evaluation can also be used to better inspect data distributions, as the higher-dimensional representations (\textit{embeddings}) that the model creates to describe the system and its atomic environments are highly informative. In order to visualize structural and atomic descriptors, we rely on Principal Component Analysis\cite{abdi2010principal} to reduce their dimensionality to the components capturing the largest variance. 
To estimate how close a structure $\mathbf{x}$ is to the training data used for fine-tuning we propose a metric based on the following score function, built on a Gaussian Kernel Density Estimate of the training distribution 
\begin{align*}
    &s(\mathbf{x}) = \log G_{KDE}(\mathbf{x}); &G_{KDE}(\mathbf{x}) = C\sum_n^N \exp\left(-\frac{\Vert P(\mathbf{x} - \mathbf{x}_n)\Vert^2}{2h^2}\right),
\end{align*}
where $C$ is the normalization constant, $P$ is the matrix projecting $\mathbf{x}$ on the first $k$ principal components of the train set, and $h$ is selected as $N^{-1/(k+4)}$.
Therefore, $s(\mathbf{x})$ represents how likely it is to find the given structure inside (or close enough) the training distribution, allowing us to mark a structure as Out Of Distribution (OOD) when its score falls below the first percentile of the training scores.
We use this metric to ask whether the coverage provided by the small periodic cells of ASSYST, generated automatically across a wide variety of crystalline phases, overlaps with bulk glass-like samples of a ternary alloy such as the one examined here, represented by the LJ data. To this end, each structure is represented by the mean, over its atoms, of the invariant node features of the pre-trained MACE-OMAT model (the parent model before fine-tuning), projected onto the first two principal components of a PCA fitted jointly on the ASSYST and LJ data. A Gaussian kernel density estimate fitted on the ASSYST structures alone then assigns to each LJ structure an OOD score, namely its log-likelihood under the training density (higher values meaning better represented); structures scoring below the first percentile of the training scores are flagged as OOD ($181$ of the $4\,037$ LJ structures, $4.5\%$). Finally, the $647$ LJ-test structures are ranked by ascending score, so that OOD rank $1$ is the structure least supported by the ASSYST data and rank $647$ the best supported. Full definitions, together with additional results of the embedding analysis (PCA maps of the ASSYST and LJ structural and atomic embeddings, which overlap to a large extent, and the full correlation between OOD rank and energy, force and stress errors), are given in Supplementary Sec.~\ref{SI-si:embeddings}.

Figure~(\ref{fig:ood_combined}) shows the force RMSE of the ASSYST-fine-tuned model (MACE-OMAT-ASSYST) on the LJ-test structures against their DFT reference energy, colored by OOD rank. The largest errors, $\approx0.32$--$0.36$~eV/\AA, roughly twice those of typical structures, belong to a small group of Expanded/Compressed glass configurations. These are at the same time the highest-energy structures of the set ($\approx-3.7$ to $-3.95$~eV/atom, against $\approx-6$~eV/atom for the bulk of the data) and the least supported by ASSYST, with OOD ranks $1$ to $4$: strongly compressed cells whose local environments lie outside the range covered by the fine-tuning data. A second Expanded/Compressed group, at $\approx-5.2$~eV/atom and with less extreme ranks, shows intermediate errors, while for the bulk of the test set the error level is set mainly by the type of configuration rather than by the OOD rank. The OOD score therefore singles out the extreme failure cases of the fine-tuned model. Such observations focused attention on the level of expansion/compression that was considered in that reference database and led to considerations on its usefulness in terms of representations and impact on performance. To this aim, a cleaned version of LJ was prepared (LJ-c), the zero-shot errors were analyzed again and the model was fine-tuned separately for comparison also on this variant. The related results are reported in Supplementary Figs.~(\ref{SI-fig:Parity_MACE_OMAT_LJ_CLEAN_3models}), (\ref{SI-fig:Performance_MACE_LJ_CLEAN_RMSE_AllPretrained}) and~(\ref{SI-fig:Performance_MACE_OMAT_LJ_CLEAN_RMSE_2models}). 

\begin{figure} [t]
    \centering
    \sbox{\panelbox}{\includegraphics[width=0.62\textwidth]{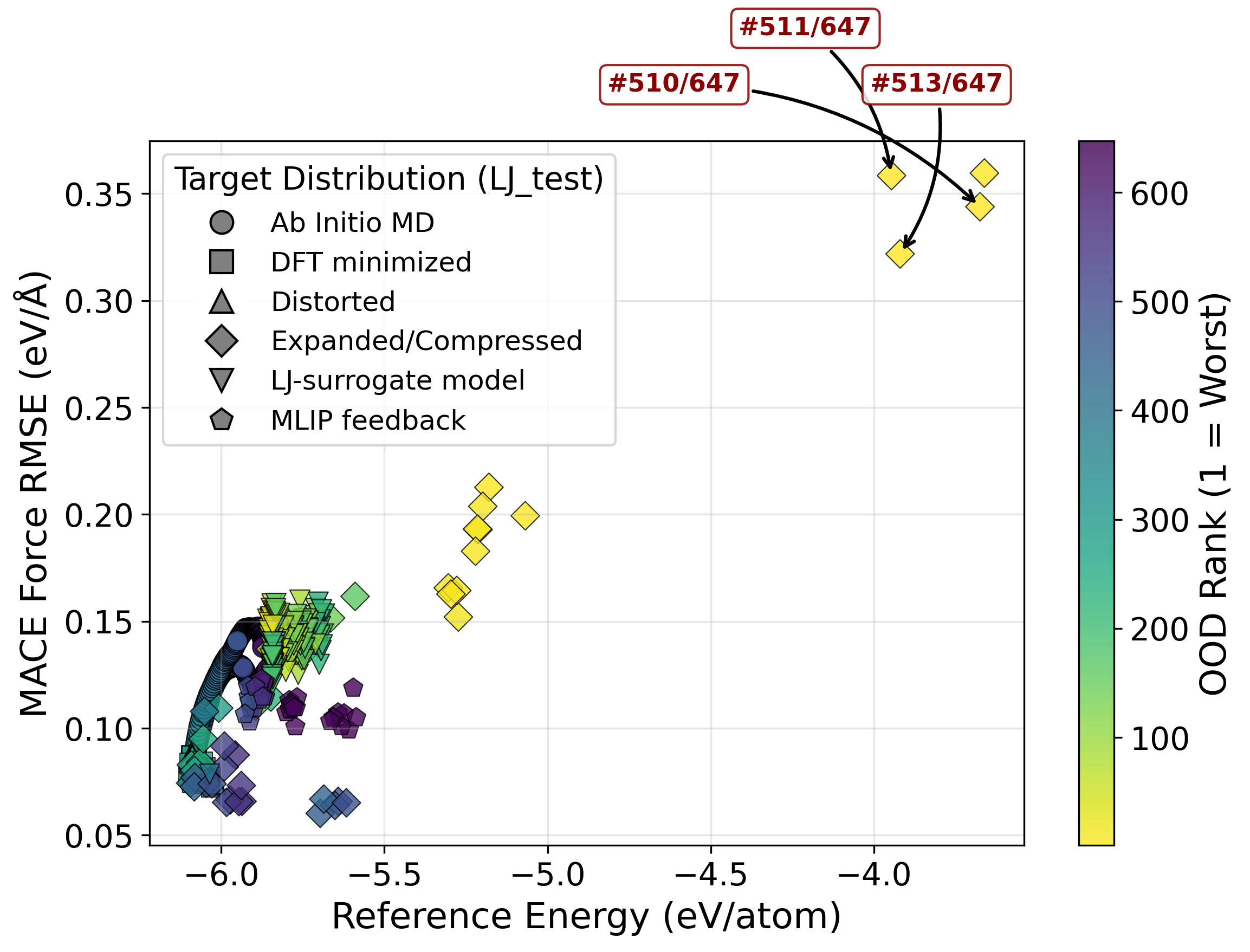}}%
    \usebox{\panelbox}\hfill
    \parbox[b][\ht\panelbox][s]{0.36\textwidth}{\centering
        \includegraphics[height=0.49\ht\panelbox]{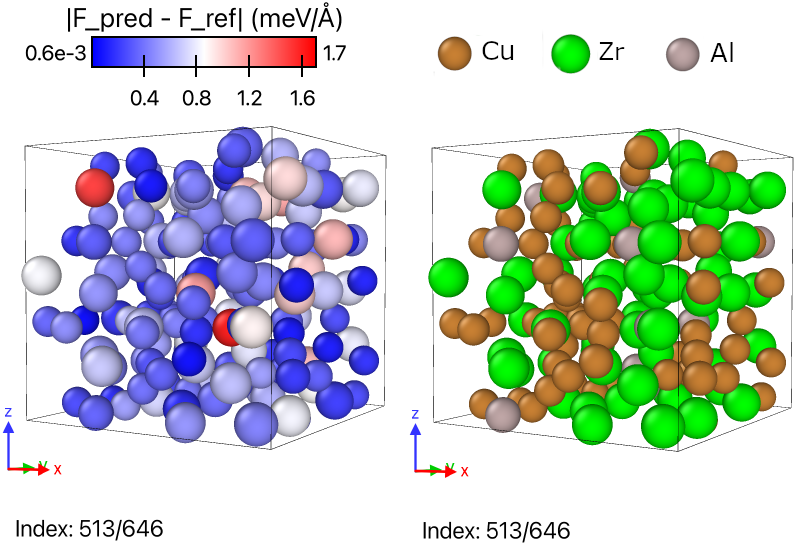}\par\vfill
        \includegraphics[height=0.49\ht\panelbox]{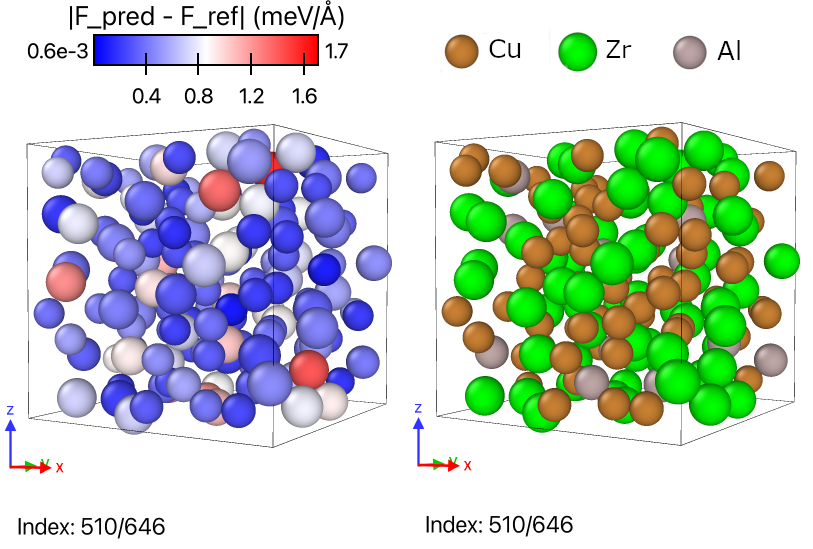}}
    \caption{Force errors of the MACE-OMAT-ASSYST model on the LJ-test set in energy space, and the configurations responsible for the largest ones. Left: per-structure force RMSE against the DFT reference energy per atom for the $647$ LJ-test structures; marker shape gives the generation class and color the out-of-distribution (OOD) rank, with rank $1$ the structure least supported by the ASSYST training data (definition in the text and in Supplementary Sec.~\ref{SI-si:embeddings}). The largest errors ($\approx0.32$--$0.36$~eV/\AA) belong to a small group of Expanded/Compressed configurations that have the lowest OOD ranks and are also the highest-energy structures of the set ($\approx-3.7$ to $-3.95$~eV/atom, against $\approx-6$~eV/atom for the bulk of the data); a second Expanded/Compressed group near $-5.2$~eV/atom shows intermediate errors ($\approx0.15$--$0.21$~eV/\AA). Labels mark structures \#510, \#511 and \#513, of OOD rank $1$, $4$ and $3$, respectively. Right: structures \#513 (top) and \#510 (bottom); in each pair, atoms are colored by the magnitude of the per-atom force error $|\mathbf{F}_\mathrm{pred}-\mathbf{F}_\mathrm{ref}|$ (left) and by species, Cu, Zr and Al (right).}
    \label{fig:ood_combined}
\end{figure}

\subsubsection{Training strategy}\label{sec:Training strategy}

Following a zero-shot performance benchmark of various foundational architectures on the LJ reference, the MACE-OMAT model was selected as the foundational backbone for downstream optimization. Transfer learning via fine-tuning (FT) enables foundation models to circumvent the high data costs of training from scratch by providing an informed parameter initialization. This optimization strategy leverages learned equivariant representations of local atomic environments, accelerating gradient convergence.
To explore model adaptation, we compare Single-Head Full-Parameter Fine-Tuning (Naive-FT) against Direct Multi-Head Fine-Tuning with Replay (MH-FT):\begin{itemize}
    \item Naive-FT: subjects the entire network weights—both the underlying message-passing layers and the terminal atomic energy prediction readouts—to a non-constrained modification against the new target coordinates. While this usually yields highly tailored results, the method exposes the network to the so-called \textit{catastrophic forgetting}, potentially corrupting previously learned generalized principles.
    \item MH-FT: a multi-head adaptation pipeline that uses simultaneous replay sampling. The deep invariant feature-extraction layers of MACE-OMAT are shared as a universal backbone across all domains, but the architecture attaches independent, low-dimensional multi-layer perceptrons (readout heads) dedicated to each dataset. Throughout the fine-tuning phase, the shared backbone parameters and the freshly initialized readout head are updated simultaneously via backpropagation. To maintain gradient stability, training batches are dynamically mixed, co-sampling structures from the target domain alongside a representative subset (replay buffer) drawn from the foundational OMAT training corpus. This multi-objective loss formulation forces the model to accommodate localized chemical variations while locking in the global constraints learned during large-scale pre-training.
\end{itemize}

\begin{figure}[t]
    \centering
    \includegraphics[width=0.99\textwidth]{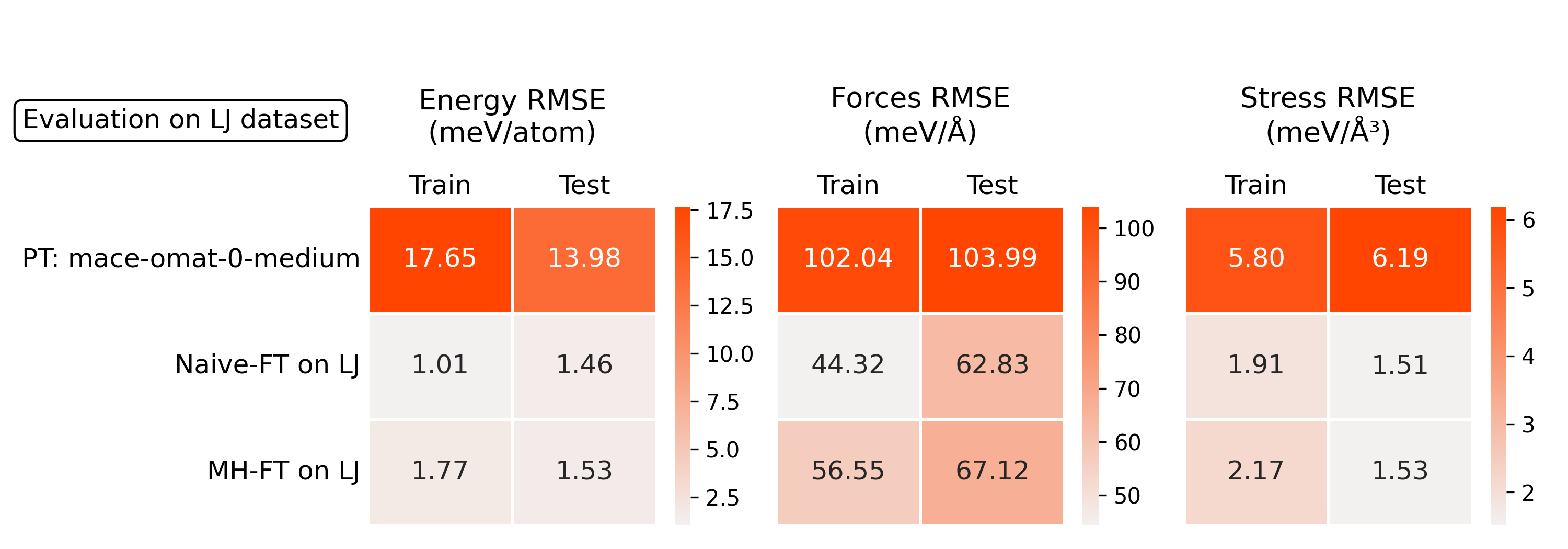}
    \caption{Evaluation of the selected pre-trained (PT) MACE-OMAT foundation model and its LJ fine-tuned (-FT) versions on the LJ-test database of CuZrAl bulk metallic glasses. The heatmaps compare the Root Mean Square Error (RMSE) for energy (meV/atom), forces (meV/\AA), and stress (meV/\AA$^3$) across LJ training and test sets.}
    \label{fig:Performance_MACE_OMAT_LJ_RMSE_3models}
\end{figure}

In Fig.~(\ref{fig:Performance_MACE_OMAT_LJ_RMSE_3models}) we show the results of the fine-tuning procedures applied to the LJ database. While the model has already seen instances of amorphous, non-glassy systems in its large pre-training step, by further updating its representations on meaningful CuZrAl-BMGs structures in the training set, the errors lower significantly. The learning curves follow the two-step procedure in which energy and force weights in the loss are changed to improve training quality and convergence. Details of all the parameters can be found in the Supplementary Tables (\ref{SI-tab:mace_hyperparameters} and \ref{SI-tab:mace_mh_ft_hyperparameters}). Further results can be seen in Supplementary Fig.~(\ref{SI-fig:naive_valid_curve_EFS}). 

The Naive-FT procedure has also been applied for fine-tuning of the MACE-OMAT model on the ASSYST (AS) database. As can be observed from Supplementary Figs.~(\ref{SI-fig:final_metrics_heatmap_no_unc}) and (\ref{SI-fig:training_dynamics_grouped}), the model adapts to the new data, with a similar learning trend to the one observed above. While for the LJ database case we decided to keep the exact train-test splitting as in the original reference work\cite{wadowski2026efficient} for comparison, on the newly generated ASSYST database, we perform a more complete analysis by varying data and model seeds, which, however, does not highlight qualitative changes in the learning process.

\subsection{About the selected EAM potential}\label{sec:EAM focus}
The database construction and training for this open-source ternary EAM potential were entirely computational, with no direct experimental parameters used during the fitting process~\cite{EAM}. The training database consisted of a potential energy surface representing over 600 configurations comprising 96-128 atoms, which included a comprehensive mix of crystalline phases, liquids, and Cu$_{46}$Zr$_{47}$Al$_{7}$ MGs simulated and evaluated using DFT. The potential parameters were fitted to the ab initio database employing the force-matching approach with the potfit program\cite{Brommer2007}. This fitting process was further optimized and refined through self-consistent recursive refinements.

\subsection{Computational procedure}\label{sec:Molecular dynamics validation} 
\subsubsection{Quenching simulations}\label{sec:Quenching simulations}
\textbf{\textit{Fixed compound composition:}} Firstly, to create the representative structures of the  $\text{Cu}_{0.46}\text{Zr}_{0.46}\text{Al}_{0.08}$ metallic glass, ten independent $6000$-atom structures were equilibrated for each interatomic potential. The equilibration was performed using the isothermal–isobaric NPT ensemble for \SI{10}{\pico\second} at \SI{2000}{\kelvin}, with the pressure held at \SI{0}{\bar}.
The next NPT ensemble quenched the samples from \SI{2000}{} to \SI{300}{K} at \SI{0}{\bar}, with $10^6$ timesteps, resulting in a cooling rate of \SI{1.7}{\kelvin\per\pico\second}. 
The timestep of both equilibration and quenching was set to \SI{0.001}{\pico\second}.\\
\textbf{\textit{Variable compound composition study:}}
To generate the samples for the composition-dependent characterization, a similar generation procedure was followed. The only difference was the system size, set to $150$ atoms, to enable DFT verification of those configurations. This is not done for the $\text{Cu}_{0.46}\text{Zr}_{0.46}\text{Al}_{0.08}$ metallic glass, as the LJ database already contains structures of this composition. The set of compositions was selected by sampling the ternary space of a supposedly good GFA region. The supporting evidence for this decision lies in two factors: literature work\cite{makinen2025bayesian} and thermodynamic arguments via inspection of the CuZrAl phase diagram at 1000~K (see Supplementary Fig.~(\ref{SI-fig:phasediag})).  
\subsubsection{Structural and elastic properties}\label{sec:Structural and elastic properties}
The structure of the $\text{Cu}_{0.46}\text{Zr}_{0.46}\text{Al}_{0.08}$ glass was quantified by calculating total and partial radial distribution functions for the $6000$-atom configurations, which were averaged over ten independent runs.

The elastic property prediction accuracy for different potentials is quantified by applying finite structural deformations to the 150-atom configurations. 
The resulting changes in potential energy and stress are used to derive the stiffness matrix components, $C_{11}$, $C_{12}$, and $C_{44}$.
This enables application of the Voigt-Reuss-Hill averaging method~\cite{hill1952elastic, singh2021mechElastic} to calculate four elastic observables---Young's modulus $E$, bulk modulus $B$, shear modulus $G$, and Poisson's ratio $\nu$.\\
\textbf{\textit{Fixed compound composition:}} In the case of $\text{Cu}_{0.46}\text{Zr}_{0.46}\text{Al}_{0.08}$ metallic glass, the distorted samples were based on the DFT-minimized configurations from the LJ-database, enabling consistent comparison with available DFT data.
To benchmark the potentials on the prediction of elastic properties (Fig.~(\ref{fig:elastic_nes_grid})), we summarize their accuracy across four elastic observables
using a single Normalized Error Score (NES) with respect to the DFT and experimental data. For a potential $m$, property $P$, cooling rate $X$ and reference $r\in\{\mathrm{DFT},\mathrm{exp}\}$ we first compute the root-mean-square error over the structures available at that cooling rate, $\mathrm{RMSE}^{\,r}_{m,P,X}$. Because the properties span very different magnitudes (e.g. $E\!\sim\!10^{2}$~GPa vs.\ $\nu\!\sim\!0.3$), each error is made dimensionless by dividing by the mean error of that property over \emph{all} potentials and cooling rates (computed separately for each reference)
$
  \varepsilon^{\,r}_{m,P,X}
  = \frac{\mathrm{RMSE}^{\,r}_{m,P,X}}
         {\big\langle \mathrm{RMSE}^{\,r}_{P}\big\rangle_{m,X}},
  \qquad
  \big\langle \mathrm{RMSE}^{\,r}_{P}\big\rangle_{m,X}
  = \frac{1}{N_m N_X}\sum_{m',X'}\mathrm{RMSE}^{\,r}_{m',P,X'} .
$
These normalized errors $\varepsilon^{\,r}_{m,P,X}$ are exactly the quantities color-coded in the grid cells of Fig.~(\ref{fig:elastic_nes_grid}). The final score of a potential is their average over the two references, the four properties and the six cooling rates,
$
  \mathrm{NES}_{m}
  = \frac{1}{2\,N_P\,N_X}
    \sum_{r}\sum_{P}\sum_{X}\varepsilon^{\,r}_{m,P,X},
$
so that $\mathrm{NES}\!\approx\!1$ denotes a potential of average accuracy across the task, while $\mathrm{NES}\!<\!1$ ($>\!1$) marks below- (above-) average error. The normalization is deliberately \emph{cross-model}: it removes the disparate physical units of the four properties while preserving the relative ranking of the potentials.\\
\textbf{\textit{Variable compound composition study:}} To verify the performance of the models across composition space, the new initial samples were generated using EAM and MACE-based interatomic potentials. The protocol described in Section \ref{sec:Quenching simulations} was the basis for the quenching procedure, but the system size was reduced to $150$ atoms. For each considered composition, we generate $10$ independent samples, which are subjected to distortions. Then, elastic properties are derived, as done for the $\text{Cu}_{0.46}\text{Zr}_{0.46}\text{Al}_{0.08}$ samples.
\subsubsection{Stress-strain response}\label{sec:Mechanical response}
To capture mechanical response behavior, the shearing simulations were performed using the $6000$-atom $\text{Cu}_{0.46}\text{Zr}_{0.46}\text{Al}_{0.08}$ configurations, representative of the temperature of $300$ K. However, to capture the mechanical response of metallic glasses, the sample subjected to shearing should originate from relatively deep regions of the potential energy landscape \cite{kaskosz2026transmutation}. Therefore, the initial samples originated from a procedure described in Section \ref{sec:Quenching simulations}, but with longer quenching of $1.5\cdot10^8$ timesteps, and using a relatively fast EAM potential. The resulting cooling rate was equal to \SI{0.0113}{\kelvin\per\pico\second}, much lower than the original \SI{1.7}{\kelvin\per\pico\second}.
Such configurations served as a starting point for shearing with EAM and MACE-based potentials. In the latter case, the quenched samples were additionally equilibrated via a short NPT run at 300 K and 0 bar for $5.0\cdot10^4$ timesteps of \SI{0.001}{\pico\second}, to compensate for the atomic force mismatch between EAM and MACE-like potentials.
In this way, for each interatomic potential, six independent samples are prepared and subsequently sheared. Mechanical deformation is performed by incrementally tilting the simulation box in the xy plane with a strain size of $0.0001$. After each step, the atomic positions are minimized via conjugate gradient method, and the simulation stops when the strain reaches a value of $0.2$.
\subsubsection{Specific heat}\label{sec:Specific heat}
The constant-pressure heat capacity was obtained by stepwise heating of eight independently quenched Cu$_{46}$Zr$_{46}$Al$_{8}$ glass samples ($N=6000$ atoms each), using a MACE machine-learning interatomic potential fine-tuned for this system and propagated in the isothermal--isobaric ensemble with a Langevin thermostat and barostat at ambient pressure (\SI{1}{\femto\second} timestep). After equilibration at \SI{300}{\kelvin} --- with the amorphous, dynamically arrested character of each sample verified through its mean-squared displacement and pair distribution function --- the samples were heated along a grid of $13$ temperatures spanning \SIrange{300}{1200}{\kelvin}, refined across the transition region. Each temperature step comprised a rapid linear ramp (\SI{50}{\kelvin\per\pico\second}), an isothermal hold, and a long isothermal production segment
(\SIrange{179}{899}{\pico\second}, longest near and above the transition) during which the enthalpy was sampled every \SI{0.01}{\pico\second}, giving $\approx\SI{5.6}{\nano\second}$ per replica in total. Every plateau was given the same \SI{25}{\pico\second} of effective equilibration before sampling began, a value verified sufficient by requiring the cross-replica coherence of the initial enthalpy drift to be statistically insignificant. Measuring under isothermal conditions rather than during a continuous scan ensures that the ensemble underlying the fluctuation formula is well defined; the full microstate --- including the barostat degrees of freedom --- was checkpointed at every plateau, so that sampling could be extended at selected temperatures as exact continuations of the same trajectory. The heat capacity was determined by two independent routes: the equilibrium enthalpy fluctuations at each plateau, $C_p = \langle\delta H^2\rangle/k_{\mathrm B}T^2$, with the kinetic contribution treated analytically (see Supplementary Eq.~\eqref{SI-eq:si-cp-split}); and the calorimetric slope $C_p = (\partial\langle H\rangle/\partial T)_P$. For a glass, these carry different physical content --- the fluctuation route probes only the degrees of freedom accessible within the observation window at fixed temperature, whereas the slope route additionally absorbs irreversible relaxation along the heating path --- and the former is therefore used as the primary measure of the calorimetric signature. Because the fluctuation formula is an equilibrium identity, the analysis window at each state point was fixed by a stationarity criterion applied to the enthalpy record itself, with no parameter adjusted to reconcile the two estimators. Particle-swap moves were deliberately omitted, as they would artificially equilibrate the configurational degrees of freedom whose freezing constitutes the transition. Fitting the plateau relaxation to saturating and non-saturating laws identifies which temperatures genuinely equilibrate and which merely age, and the baselines defining the heat-capacity step were restricted to the former. The glass transition temperature was extracted from the heat-capacity step, from the kinks in $\langle U\rangle(T)$ and $\langle V\rangle(T)$, and from the onset of structural relaxation; uncertainties combine batch-means statistics within each trajectory, the dispersion across the eight independent replicas, and the choice of baseline windows. The same protocol was applied to a second interatomic potential --- each acting on glasses produced by its own quench --- and, by comparison against continuous ramps spanning nearly two orders of magnitude in heating rate, was used to calibrate how slowly a direct scan must be driven to recover equilibrium thermodynamics. Full details are given in the Supplementary Methods (Sec.~\ref{SI-si:cp-methods}).


\section*{\textcolor{black}{Data Availability statement}}
The data that support the findings of this study, including the generated datasets, the trained models and the simulation outputs, are available from the corresponding author upon reasonable request.

\section*{Acknowledgments}

We gratefully acknowledge Polish high-performance computing infrastructure PLGrid (HPC Center: ACK Cyfronet AGH) for providing computer facilities and support within computational grant no. PLG/2025/018702. 

D.M., K.C. and P.S. acknowledge support from the Polish Ministry of Science and Higher Education (MNiSW) through the targeted grant for the ``AI x Science Connect Poland'' initiative (AIxSCP), agreement No.~MNiSW/2026/DIR/1121.

L.L. and C.F. acknowledge support by the  National Recovery and Resilience Plan (NRRP), Mission 4 Component 2 Investment 1.3 - Project NEST (Network 4 Energy Sustainable Transition) and CN-HPC grant no. (CUP) J33C22001170001, SPOKE 7, of Ministero dell’Università e della Ricerca (MUR), funded by the European Union – NextGenerationEU. This research was partially funded by the Austrian Science Fund (FWF) 10.55776/F8100 project TACO.

A.W. and S.B. are supported by the European Union Horizon 2020 research and innovation program under grant agreement no. 857470 and from the European Regional Development Fund via the Foundation for Polish Science International Research Agenda PLUS program grant No. MAB PLUS/2018/8. S.B. acknowledges support from the National Science Center in Poland through the SONATA BIS grant DEC-2023/50/E/ST3/00569 (PI: S.B.). S.B. and A.D.S.P. are supported by the project ``Projektowanie Ulepszonych Szkieł Metalicznych'' (FIRST TEAM FENG.02.02-IP.05-0177/23, PI: S.B.) under the 2.2 First Team program of the Foundation for Polish Science co-financed by the European Union from the European Funds for Smart Economy 2021–2027 (FENG).

C.A. was supported by the ``MagTop'' project (FENG.02.01-IP.05-0028/23) carried out within the ``International Research Agendas'' program of the Foundation for Polish Science, co-financed by the
European Union under the European Funds for Smart Economy 2021-2027 (FENG). 

J.W. acknowledges support from the statutory funds of the Faculty of Materials Science and Engineering, Warsaw University of Technology.

\section*{Author Contribution}
Conceptualization: D. Massa, S. Bonfanti, P. Sankowski.\\
Methodology and calculations: D. Massa, A. Wadowski, L. Leoni, K. Cybinski.\\
Supervision: D. Massa, P. Sankowski, S. Bonfanti, J. Wrobel, A. Parmar, C. Franchini, C. Autieri.\\
Writing: All authors.\\
Discussion and reviewing: All authors.

\section*{Competing Interests}
The authors declare no competing interests.

\bibliography{bibl}

\IfFileExists{Supplementary.pdf}{%
  \clearpage
  \includepdf[pages=-,pagecommand={\thispagestyle{empty}}]{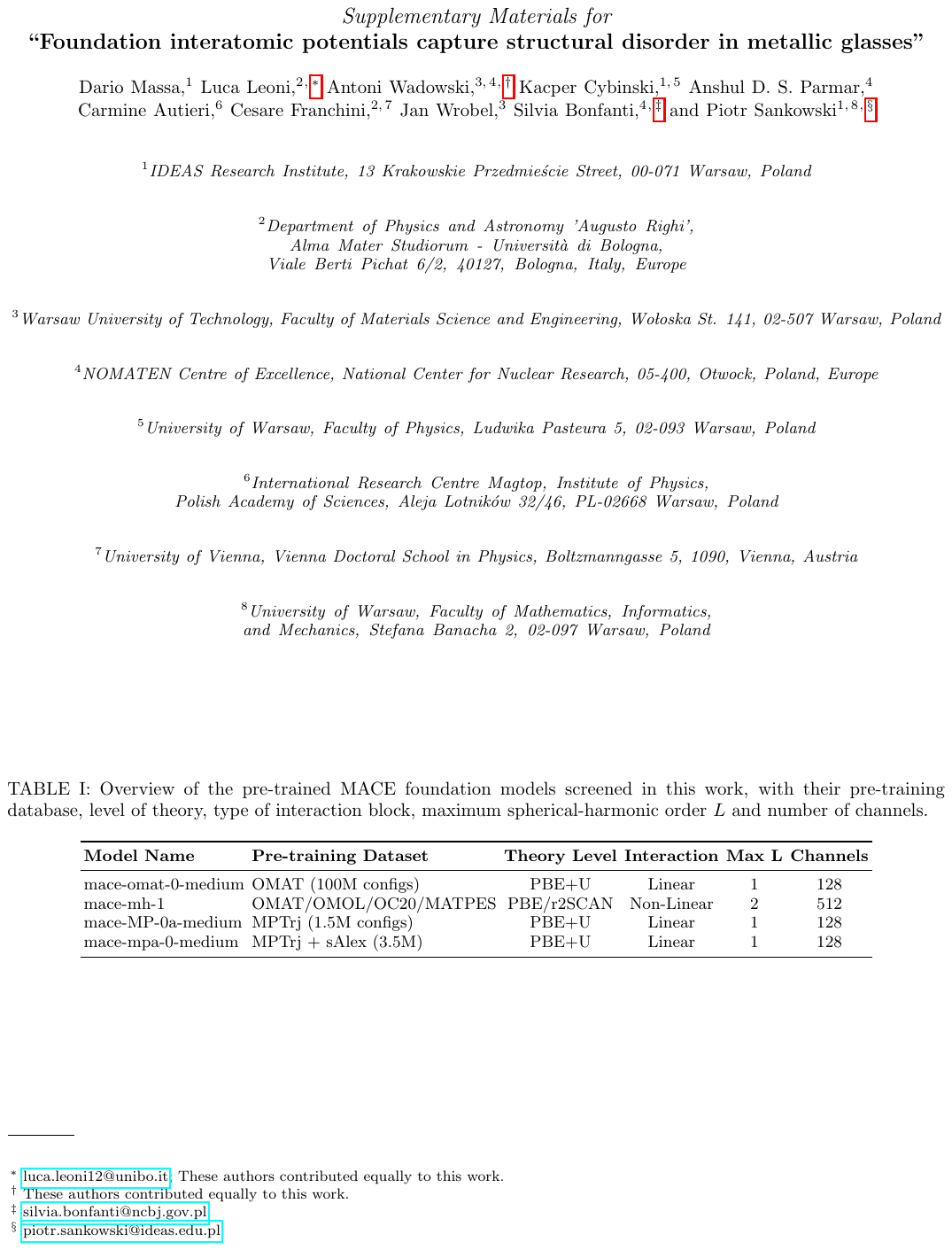}%
}{\typeout{Supplementary.pdf not found: Supplementary Material NOT appended.}}

\end{document}